\documentclass[a4paper,twocolumn,11pt,accepted=2024-02-08]{quantumarticle}
\pdfoutput=1
\usepackage[T1]{fontenc}
\usepackage{amsmath,amsfonts,amssymb,amsthm,bbm,bm, braket}
\usepackage[numbers]{natbib}
\usepackage{wasysym}
\usepackage{comment}
\usepackage{scalerel}
\usepackage{enumerate}
\usepackage{amssymb}
\usepackage{graphicx}
\usepackage{mathtools}
\usepackage{ifthen}
\usepackage{tensor}
\usepackage{tikz}
\usepackage{tikz-network}
\usetikzlibrary{patterns,decorations.pathreplacing}

\usepackage{color}
\usepackage{longtable}
\usepackage[normalem]{ulem} 

\theoremstyle{break}        
\usepackage{color}
\definecolor{myred}{RGB}{232,102,102}
\definecolor{myblue}{RGB}{187,187,255}
\definecolor{myorange0}{RGB}{252,226,5}
\definecolor{myorange0c}{RGB}{255,255,255}
\definecolor{myorange}{RGB}{255,165,0}
\definecolor{mygrey}{RGB}{105,105,105}
\definecolor{OliveGreen}{RGB}{85,107,47}
\definecolor{NavyBlue}{RGB}{0,0,128}
\definecolor{mygreen}{RGB}{34,139,34}
\definecolor{myY}{RGB}{220,255,203}
\definecolor{myYO}{RGB}{255, 220, 151}

\definecolor{mygreenc}{RGB}{150,50,50}

\usepackage{tensor}
\newcommand{\be}{\begin{equation}}
\newcommand{\ee}{\end{equation}}
\newcommand{\ba}{\begin{aligned}}
\newcommand{\ea}{\end{aligned}}
\newcommand{\bw}{\begin{widetext}}
\newcommand{\ew}{\end{widetext}}
\newcommand{\1}{\mathbbm{1}}

\theoremstyle{plain}

\theoremstyle{plain}

\theoremstyle{plain}

\usepackage[colorlinks,bookmarks=false,citecolor=NavyBlue,linkcolor=OliveGreen,urlcolor=blue]{hyperref}

\newcommand{\Wgatedagger}[2]{
\draw[very thick] (#1-0.5, #2 +0.5) -- (#1+0.5,#2-0.5);
\draw[very thick] (#1-0.5,#2-0.5) -- (#1+0.5,#2+0.5);
\draw[ thick, fill=mygreenc, rounded corners=2pt] (#1-0.25,#2+0.25) rectangle (#1+0.25,#2-0.25);
\draw[thick] (#1,#2+0.15) -- (#1+0.15,#2+0.15) -- (#1+0.15,#2);
}
\newcommand{\Wgatered}[2]{
\draw[very thick] (#1-0.5, #2 +0.5) -- (#1+0.5,#2-0.5);
\draw[very thick] (#1-0.5,#2-0.5) -- (#1+0.5,#2+0.5);
\draw[ thick, fill=myred, rounded corners=2pt] (#1-0.25,#2+0.25) rectangle (#1+0.25,#2-0.25);
\draw[thick] (#1,#2+0.15) -- (#1+0.15,#2+0.15) -- (#1+0.15,#2);
}
\newcommand{\Wgateblue}[2]{
\draw[very thick] (#1-0.5, #2 +0.5) -- (#1+0.5,#2-0.5);
\draw[very thick] (#1-0.5,#2-0.5) -- (#1+0.5,#2+0.5);
\draw[ thick, fill=myblue, rounded corners=2pt] (#1-0.25,#2+0.25) rectangle (#1+0.25,#2-0.25);
\draw[thick] (#1,#2+0.15) -- (#1+0.15,#2+0.15) -- (#1+0.15,#2);
}

\newcommand{\WgateblueT}[2]{
\draw[very thick] (#1-0.5, #2 +0.5) -- (#1+0.5,#2-0.5);
\draw[very thick] (#1-0.5,#2-0.5) -- (#1+0.5,#2+0.5);
\draw[ thick, fill=myblue, rounded corners=2pt] (#1-0.25,#2+0.25) rectangle (#1+0.25,#2-0.25);
\draw[thick] (#1,#2-0.15) -- (#1+0.15,#2-0.15) -- (#1+0.15,#2);
}
\newcommand{\Wgategreen}[2]{
\draw[very thick] (#1-0.5, #2 +0.5) -- (#1+0.5,#2-0.5);
\draw[very thick] (#1-0.5,#2-0.5) -- (#1+0.5,#2+0.5);
\draw[ thick, fill=mygreen, rounded corners=2pt] (#1-0.25,#2+0.25) rectangle (#1+0.25,#2-0.25);
\draw[thick] (#1,#2+0.15) -- (#1+0.15,#2+0.15) -- (#1+0.15,#2);
}
\newcommand{\MYcircle}[2]{
\draw[thick, fill=white] (#1,#2) circle (0.1cm); }
\newcommand{\MYsquare}[2]{
 \coordinate (Origin) at (#1,#2);
\filldraw [thick, fill=white, even odd rule] ($(Origin)+(-.1cm,-.1cm)$) coordinate (Square) -- ++(0.0cm,0.2cm) -- ++(0.2cm,0.0cm) -- ++(0.0cm,-0.2cm) -- cycle;
 }
\newcommand{\MYtriangle}[2]{
 \coordinate (Origin) at (#1,#2);
\filldraw [thick, fill=white, even odd rule] ($(Origin)+(-.0cm,{0.666*cos(60)*0.3cm})$) coordinate (Triangle) -- ++(0.15cm,-{cos(60)*0.3cm}) -- ++(-0.3cm,0.0cm) -- ++(0.15cm,{cos(60)*0.3cm}) -- cycle;
}
\newcommand{\MYcircleB}[2]{
\draw[thick, fill=black] (#1,#2) circle (0.1cm); }

\newcommand{\rhoO}[2]{
\draw[very thick] (-.5+#1,0.5+#2) -- (#1,0+#2);
\draw[very thick] (#1,0+#2) -- (.5+#1,0.5+#2);
\draw[very thick] (-.5+#1,#2) -- (.5+#1,#2);
\draw[ thick, fill=mygreen, rounded corners=2pt] (-0.35+#1,0.2-0.25+#2) rectangle (0.35+#1,0.2+0.2+#2);
\draw[very thick] (0.1+#1,0.15+.18+#2)-- (.15+0.1+#1,0.15+.18+#2) -- (.15+0.1+#1,0+.18+#2);
}

\newcommand{\mcirc}{\mathbin{\scalerel*{\fullmoon}{G}}}

\newcommand{\mcircf}{\mathbin{\scalerel*{\newmoon}{G}}}

\newcommand{\MYsquareB}[2]{

\draw[ thick, fill=black, rounded corners=2pt] (#1-0.25,#2+0.25) rectangle (#1+0.25,#2-0.25);
\draw[thick, color=white] (#1,#2+0.15) -- (#1+0.15,#2+0.15) -- (#1+0.15,#2);
}

\definecolor{skyblue}{RGB}{135, 206, 235}
\hypersetup{
 pdftitle={Hierarchical generalization of dual unitarity},
 pdfsubject={Many-body systems},
 pdfauthor={Xie-Hang Yu, Xie-Hang Yu, and Pavel Kos},
 pdfkeywords={exact solutions, quantum circuits, quantum many-body systems}
}

\begin{document}

\title{Hierarchical generalization of dual unitarity}

\author{Xie-Hang Yu}
\author{Zhiyuan Wang}
\author{Pavel Kos}%

\affiliation{Max-Planck-Institut f{\"{u}}r Quantenoptik, Hans-Kopfermann-Str. 1, 85748 Garching, Germany}%


\begin{abstract}
Quantum dynamics with local interactions in lattice models display rich physics, but is notoriously hard to study. Dual-unitary circuits allow for exact answers to interesting physical questions in clean or disordered one- and higher-dimensional quantum systems. 
However, this family of models shows some non-universal features, like vanishing correlations inside the light-cone and instantaneous thermalization of local observables.
In this work we propose a generalization of dual-unitary circuits where the exactly calculable spatial-temporal correlation functions display richer behavior, and have non-trivial thermalization of local observables. This is achieved by generalizing the 
single-gate condition to a hierarchy of multi-gate conditions, where the first level recovers dual-unitary models, and the second level exhibits these new interesting features. 
We also extend the discussion and provide exact solutions to correlators with few-site observables and discuss higher-orders, including the ones after a quantum quench. In addition, we provide exhaustive parametrizations for qubit cases, and propose a new family of models for local dimensions larger than two, which also provides a new family of dual-unitary models.
%
%
\end{abstract}
\maketitle

\section{Introduction} \label{sec:Intro}

One of the pivotal problems in quantum many-body physics is understanding the dynamics of extended systems with local interactions. Although the local interactions are simple, they generate complex dynamics, which is in general too hard to describe.
The dynamics can be characterized with different probes such as local correlation functions, entanglement spreading, and other quantum information quantities.
The understanding of this type of dynamics is currently at the center of attention in many fields spanning from nonequilibrium statistical mechanics, quantum information, condensed matter to high-energy physics and quantum gravity.


However, the complexity of quantum dynamics, both analytically and numerically, presents a significant hurdle. For example, the bond dimension of Matrix Product States (MPS) typically increases exponentially due to the linear growth in entanglement entropy~\cite{Daley2004time,Schuch2008entropy}. 
This necessitates the use of solvable models to unravel many-body behavior. The most well-known examples are noninteracting (Gaussian) systems such as free fermions or bosons, Clifford circuits, and Bethe-ansatz interacting integrable models~\cite{PhysRevLett.122.150605}. 
Unfortunately, all of these models are not chaotic, in contrast to generic examples. 
If one is prepared to average over an ensemble of systems, random unitary circuits~\cite{
fisher2023random} provide examples of solvable chaotic dynamics. However, the averaged results miss a lot of relevant physics and are less relevant for translationally invariant systems.


A recently discovered solvable family of models, known as \emph{dual-unitary}  quantum circuits~\cite{bertini2019exact}, has distinctly different solvable structure which does not require averaging, and moreover contains both integrable and chaotic examples.
The basic property, which enables solvability is the unitarity of the local gates in the space direction. In this paper, we generalize this to a condition on a few gates and unravel new families of solvable models.

Dual unitarity was shown to enable analytical computations of correlation functions~\cite{bertini2019exact,piroli2020exact},
chaos indicator spectral form factor~\cite{bertini2018exact,bertini2021random}, operator and entanglement spreading~\cite{piroli2020exact, bertini2019entanglement,bertini2020operator, gopalakrishnan2019unitary,claeys2020maximum,bertini2020scrambling, reid2021entanglement, zhou2022maximal}, 
deep thermalization through emergent state designs~\cite{ho2022exact,claeys2022emergent,ippoliti2022dynamical}, study of eigenstate thermalization~\cite{fritzsch2021eigenstate} and
temporal entanglement~\cite{lerose2020influence,Giudice2022temporal, foligno2023temporal}.
They also proved useful in connections with measurement induced phase transitions~\cite{ippoliti2021postselection, ippoliti2021fractal,lu2021spacetime}, had aspects of their computational power characterized~\cite{suzuki2021computational}, and have already been realized in experimental setups~\cite{chertkov2021holographic, mi2021information}. 
The exhaustive parametrization is simple and is known for dual-unitary gates of two qubits~\cite{bertini2019exact}.
In general, however, only certain non-exhaustive families of gates are known~\cite{rather2020creating, gutkin2020exact, claeys2021ergodic, aravinda2021from,prosen2021manybody, marton2022construction, mestyan2022multidirectional,claeys2023dualunitary}.
Some extension of the dual-unitary condition were already proposed. One can add arbitrary perturbations and perform a perturbative analysis~\cite{kos2021correlations, rampp2023from}, generalize to the case of having three or more unitary directions~\cite{jonay2021triunitary, ternary2022milbradt, mestyan2022multidirectional}, study random and hologoraphic geometries~\cite{kasim2023dual,masanes2023discrete}, and lift the ideas to open systems \cite{kos2023circuits} and classical symplectic circuits~\cite{christopoulos2023dual}.

Despite the success of the dual-unitary circuits (and its extension) in describing  physical properties of non-equibrium dynamics, they exhibit some non-universal features. 
Firstly, in dual-unitary circuits the non-vanishing correlation functions exist only on light-cones edges. 
Secondly, the thermalization of local observables is instantaneous. The fundamental reason behind these non-universal features is that their single gate's conditions are too restrictive. In fact, this family of models doesn't even include many gates that are solvable by other methods, e.g. the Identity, Controlled-Not and Controlled-Z.
This raises the question: Can the dual-unitary condition be relaxed to allow for richer physics while still maintaining the solvability of the spatial-temporary correlation function? 


In this paper, we answer these questions in the affirmative by relaxing the dual-unitary condition and extending it 
to a hierarchy of conditions that contains more and more local gates.
The dual-unitary condition on one gate forms the the first level of the Hierarchy denoted by $\mathfrak{L}_1$, whereas the circuit with two-gate condition at the second level of the Hierarchy $\mathfrak{L}_2$ allows for the exact calculation of two-point correlation functions, which are richer than for dual unitaries, i.e. non-vanishing at the same site and different times.
Moreover, when quenched from the solvable initial states, the $\mathfrak{L}_2$ circuit exhibits non-trivial thermalization of local observables. 
We go beyond $\mathfrak{L}_2$ and show that higher levels of hierarchical circuits limit the maximum speed of information spreading. 

We provide complete parameterization for the $\mathfrak{L}_2$ and $\mathfrak{L}_3$ circuits in the qubit case. For the larger local Hilbert space dimensions, we propose a new method to construct a class of circuits using the Clifford group that is analytically trackable. This method can also be used to construct new families of dual-unitary circuit. 

The paper is structured as follows: In Sec. \ref{sec:Hierarchical}, we introduce the notation. Subsec. \ref{sec:DU} reviews the dual-unitary circuits and introduces our parametrization method using Clifford groups. The hierarchical generalization is outlined in subsec. \ref{sec:HG}. After that, we dive into details of the $\mathfrak{L}_2$ and $\mathfrak{L}_3$ circuits in subsecs. \ref{subsec:Second-Hierarchy} and \ref{subsec:Third-Hierarchy}, including their parametrization. In Sec. \ref{sec:applications}, we discuss the physical applications of the different levels of Hierarchical circuits. Subsec. \ref{subsec:STCF} considers the two-point correlation functions for $\mathfrak{L}_1$, $\mathfrak{L}_2$ and $\mathfrak{L}_3$ circuits. In subsec. \ref{subsec:Biggersupports}, we extend our discussion to the correlation functions of multisite observables, and three-point correlation functions. Subsec. \ref{Quantum_quench} discusses the evolution of an initial state from a quantum quench. We generalize the solvable initial states~\cite{piroli2020exact} for the $\mathfrak{L}_2$ circuit and explore the relationship between quench dynamics and quantum thermalization. In Sec. \ref{sec:conclusions}, we summarize the main results of the paper and discuss future directions.

\section{Hierarchical generalization of dual unitarity}\label{sec:Hierarchical}

In this paper we consider a chain comprised of $L$ cells, with each cell containing $2$ sites at integers and odd-half integer sites. At each site there is a Hilbert space with a local dimension $D$. Consequently, the corresponding
total Hilbert space is $\mathcal{H}=(C^{D})^{2L}$. The local basis is denoted by $\ket{j}$ with $j=0,1,\cdots,D-1$. The chain's dynamics is governed by a brickwall Floquet circuit
\begin{equation}
\begin{aligned}
\mathbb{U}&=\mathbb{T}_{2L}u^{\otimes L}\mathbb{T}_{2L}^{\dagger}u^{\otimes L}\\
&=
\begin{tikzpicture}[baseline=(current  bounding  box.center), scale=0.55]
\foreach \i in {0,...,4}
{
\Wgatered{2*\i}{-3+0.5};
}
\foreach \i in {0,...,2}
{
\Wgatered{2*\i+1}{-1-0.5}
}
\foreach \i in {0,1,2,3}
{
\Text[x=2*\i-0.5,y=-3.5]{\small$\i$};
}
\foreach \i in {1,3,5}
{
\Text[x=\i-0.5,y=-3.5]{\small$\frac{\i}{2}$}
}
\Text[x=8.5,y=-3.5]{\small$L$};
\Text[x=7,y=-3.5]{\small$\cdots$};

\foreach \j in {0,1}
{
\Text[x=9,y=2*\j-3]{\small$\j$}
}
\foreach \j in {1}
{
\Text[x=9,y=\j-3]{\small$\frac{\j}{2}$}
}
\foreach \i in {0,1,2,3,4,5,9}
{
\draw[gray, dashed] (\i-0.5,-0.5) -- (\i-0.5,-3);
}
\foreach \j in {0,1,2}
{
\draw[gray, dashed] (-0.5,\j-3) -- (6.75,\j-3);
\draw[gray, dashed] (7.5,\j-3) -- (8.5,\j-3);
}
\Text[x=7,y=-2.5]{\small$\cdots$};
\Text[x=7,y=-1.5]{\small$\cdots$};
\end{tikzpicture}
\end{aligned}
,
\end{equation}
where $\mathbb{T}_{2L}$ is a periodic translation operator on $2L$ sites, and $u$ a local gate. Here, for simplicity, we assume translational invariance of the circuit and introduce periodic boundary conditions. However, our result can be easily generalized to non-uniform cases and open boundary conditions.
Above, we graphically represented local unitary gates with dimension
$D^{2}\times D^{2}$ by a box with incoming and outgoing legs, \begin{equation}
u=\begin{tikzpicture}[baseline=(current  bounding  box.center), scale=0.55]
\Wgatered{0}{0}
\end{tikzpicture}
,
\qquad
\qquad
u^{\dagger}=\begin{tikzpicture}[baseline=(current  bounding  box.center), scale=0.55]
\Wgateblue{5}{5}
\end{tikzpicture},
\end{equation}
satisfying unitarity conditions
\begin{equation}\label{eq:unitarity}
\begin{aligned}
&uu^{\dagger}=
\begin{tikzpicture}[baseline=(current  bounding  box.center), scale=0.55]
\Wgatered{0}{0}
\Wgateblue{0}{-1}
\Text[x=1.5,y=-0.5]{$=$}
\draw[very thick] (2.5,0.5)--(2.5,-1.5);
\draw[very thick] (3.5,0.5)--(3.5,-1.5);
\end{tikzpicture}
=I_{D^2},\\
&u^{\dagger}u=
\begin{tikzpicture}[baseline=(current  bounding  box.center), scale=0.55]
\Wgateblue{0}{0}
\Wgatered{0}{-1}
\Text[x=1.5,y=-0.5]{$=$}
\draw[very thick] (2.5,0.5)--(2.5,-1.5);
\draw[very thick] (3.5,0.5)--(3.5,-1.5);
\end{tikzpicture}
=
I_{D^2}.
\end{aligned}
\end{equation}

Our results can be more succinctly expressed in the folded picture, where an operator over $(C^D)^{2L}$ is vectorized to a vector in $(C^D)^{4L}$ by the linear map on the basis
\begin{equation}
\ket{m}\bra{n}\to\ket{m}\ket{n}.
\end{equation}
The time evolution in Schrodinger picture can also be vectorized to
\begin{equation}
u()u^{\dagger}\to u\otimes u^*.
\end{equation}
Graphically, $u^{\dagger}$
is folded back behind $u$, thereby forming a joint operator $w 	\equiv u\otimes u^{*}$.
It is also convenient to denote the vectorized identity operator in the folded picture as an empty bullet $\ket{\mcirc}=\frac{1}{\sqrt{D}}\ket{I_D}$, which is shown below 
\begin{equation}
w=\begin{tikzpicture}[baseline=(current  bounding  box.center), scale=0.7]
\Wgategreen{0}{0}
\end{tikzpicture}
=
\begin{tikzpicture}[baseline=(current  bounding  box.center), scale=0.7]
\WgateblueT{0.15}{0.07}
\Wgatered{-0.1}{-0.07}
\end{tikzpicture}
,
\qquad
\qquad
\begin{tikzpicture}[baseline=(current  bounding  box.center), scale=0.7]
\MYcircle{0.25}{0.25}
\draw[very thick] (-0.25,-0.25)--(0.20,0.20);
\end{tikzpicture}
=\frac{1}{\sqrt{D}}
\begin{tikzpicture}[baseline=(current  bounding  box.center), scale=0.7]
\draw[very thick] (-0.25,-0.25)--(0.25,0.25)--(0.05,0.45)--(-0.45,-0.05);
\end{tikzpicture}
.
\end{equation} 
With these notations, the unitarity condition~\eqref{eq:unitarity} is graphically expressed as $
\begin{tikzpicture}[baseline=(current  bounding  box.center), scale=0.55]
\Wgategreen{0}{0}
\MYcircle{0.55}{0.55}
\MYcircle{-0.55}{0.55}
\end{tikzpicture}
=
\begin{tikzpicture}[baseline=(current  bounding  box.center), scale=0.55]
\draw[very thick] (-0.55,-0.5) -- (-0.55,0.5);
\draw[very thick] (0.55,-0.5) -- (0.55,0.5);
\MYcircle{0.55}{0.55}
\MYcircle{-0.55}{0.55}
\end{tikzpicture}
$ and $
\begin{tikzpicture}[baseline=(current  bounding  box.center), scale=0.55]
\Wgategreen{0}{0}
\MYcircle{0.55}{-0.55}
\MYcircle{-0.55}{-0.55}
\end{tikzpicture}
=
\begin{tikzpicture}[baseline=(current  bounding  box.center), scale=0.55]
\MYcircle{0.55}{-0.55}
\MYcircle{-0.55}{-0.55}
\draw[very thick] (-0.55,-0.5) -- (-0.55,0.5);
\draw[very thick] (0.55,-0.5) -- (0.55,0.5);
\end{tikzpicture}
$.

\subsection{Dual Unitarity}\label{sec:DU}

Understanding the dynamics of extended locally interacting systems is at the core of quantum many-body physics. However, this problem is usually analytically intractable and numerically exponentially hard. 
To make progress, we need some additional structure. One possibility is to demand the so-called dual-unitarity condition~\cite{bertini2019exact} mentioned in the introduction, which enables various exact calculations even for chaotic dynamics.

Dual unitarity demands that the gate $u$ is unitary even if we exchange the roles of space and time. This switching corresponds to changing which are input and output legs of the gate, resulting in the dual local gate $\tilde{u}$.
It is formally introduced by reshuffling the
indices
\be
\label{eq:tildeqgate}
\tilde{u}=\begin{tikzpicture}[baseline=(current  bounding  box.center), scale=.7]
\draw[ thick] (-4.25,0.5) -- (-3.25,-0.5);
\draw[thick] (-4,-0.25)--(-4.5,-0.75);
\draw[thick] (-4,0.25)--(-4.25,0.5)--(-4.25,-0.75);
\draw[thick] (-3.5,0.25)--(-3,0.75);
\draw[thick] (-3.5,-0.25)--(-3.25,-0.5)--(-3.25,0.75);
\draw[ thick, fill=myred, rounded corners=2pt] (-4,0.25) rectangle (-3.5,-0.25);
\draw[thick] (-3.75,0.15) -- (-3.75+0.15,0.15) -- (-3.75+0.15,0);
\Text[x=-4.25,y=-0.75]{}
\end{tikzpicture}\;, \, \,
\bra{j}\bra{l}\tilde{u}\ket{i}\ket{k}=\bra{k}\bra{l}u\ket{i}\ket{j}.
\ee
A gate is dual-unitary~\cite{bertini2019exact} if both $u$ and $\tilde{u}$ are unitary, so in addition to ~\eqref{eq:unitarity} we also require 
\begin{equation}
\tilde{u}^{\dagger}\tilde{u}=\tilde{u}\tilde{u}^{\dagger}=I_{D^{2}}\label{eq:algebric_dual_unitary},
\end{equation}
which in the folded graphical language yields
\begin{equation}
\begin{tikzpicture}[baseline=(current  bounding  box.center), scale=0.55]
\Wgategreen{0}{0}
\MYcircle{0.55}{0.55}
\MYcircle{0.55}{-0.55}
\end{tikzpicture}
=
\begin{tikzpicture}[baseline=(current  bounding  box.center), scale=0.55]
\MYcircle{0.55}{0.55}
\MYcircle{0.55}{-0.55}
\draw[very thick] (-0.5,0.55) -- (0.5,0.55);
\draw[very thick] (-0.5,-0.55) -- (0.5,-0.55);
\end{tikzpicture}
,
\qquad
\begin{tikzpicture}[baseline=(current  bounding  box.center), scale=0.55]
\Wgategreen{0}{0}
\MYcircle{-0.55}{0.55}
\MYcircle{-0.55}{-0.55}
\end{tikzpicture}
=
\begin{tikzpicture}[baseline=(current  bounding  box.center), scale=0.55]
\MYcircle{-0.55}{0.55}
\MYcircle{-0.55}{-0.55}
\draw[very thick] (-0.5,0.55) -- (0.5,0.55);
\draw[very thick] (-0.5,-0.55) -- (0.5,-0.55);
\end{tikzpicture}
. \label{eq:figure_dual_unitary_condi}
\end{equation}

The family of models defined in this way encompasses free, interacting integrable and chaotic models~\cite{bertini2019exact}. The parametrization of dual-unitary gates for $D=2$
has been fully determined~\cite{bertini2019exact}.
Despite a lack of the complete parametrization of dual-unitary gates for $D\geq3$, several families have been proposed~\cite{rather2020creating, gutkin2020exact, claeys2021ergodic, aravinda2021from,prosen2021manybody, marton2022construction}. 
We proceed to add another family to the list
, resulting in a novel extensive family of dual-unitary gates in higher dimensions.

\subsection{New Parametrization of Qudit Gates}

In the following we provide a non-exhuasive parameterization of qudit gates, which will prove useful for constructing examples at $D>2$, both for dual-unitary gates and their generalizations. 
This parametrization first appeared in Ref.~\cite{JonEtyson_2003} in the study of Operator-Schmidt decomposition, and is a special case of a novel, more general framework we report in App. \ref{Appendix:general}; however, since we do not need this more general framework in the following analysis, we restrict ourselves to the special case in the main text. 

Consider the following family of two-qudit unitary gates:
\begin{equation}
u=(v_{1}\otimes v_{2}) \ u_0 \ (v_{3}\otimes v_{4}),
\label{eq:2quditUgeneralform}
\end{equation}
where $v_{1},v_{2},v_{3},v_{4}$ are single site unitary gates, and $u_0$ is defined as
\begin{equation}
u_0=\sum_{0\leq p,q\leq D-1} \theta_{p,q}\ket{\psi_{p,q}}\bra{\psi_{p,q}}.
\label{eq:CliffordParametrization}
\end{equation}
Here $\{\theta_{p,q}\}_{0\leq p,q\leq D-1}$ is a collection of $\mathrm{U}(1)$ phases, and $\{\ket{\psi_{p,q}}\}_{0\leq p,q\leq D-1}$ is an orthonormal basis for the 2-qudit Hilbert space $(\mathbb{C}^D)^2$ defined as
\begin{equation}
\ket{\psi_{p,q}}\equiv \frac{1}{\sqrt{D}}\sum_{0\leq i,j\leq D-1}(\tau^p\sigma^q)^*_{ij}\ket{i}\otimes\ket{j},
\label{eq:psi_pq}
\end{equation}
where $\sigma,\tau$ are the $D\times D$ dimensional generators of the Clifford group satisfying the relations
\begin{equation}\label{eq:ZnCliffordCR}
    \sigma^D=\tau^D=1,~~~ \sigma\tau=\omega\tau\sigma,
\end{equation}
with $\omega=e^{2\pi i/D}$ a $D$-th root of unity. 
The matrices $\sigma,\tau$ generate the full matrix algebra $M_D(\mathbb{C})$, and we can always choose $\sigma$ to be diagonal and $\tau$ to be real. Explicitly, they are defined as 
\begin{eqnarray}\label{eq:def_sigmatau}
    \sigma&=&\sum^{D-1}_{j=0}\omega^{j}\ket{j}\bra{j},\nonumber\\
    \tau&=&\sum^{D-1}_{j=0}\ket{j+1}\bra{j},\label{eq:definition_of_Clifford}
\end{eqnarray}
where $\ket{D}\equiv \ket{0}$.

We now investigate the conditions on the parameters $\{\theta_{p,q}\}_{0\leq p,q\leq D-1}$ for $u$ to be a dual-unitary gate. After a space-time reshuffling of the indices defined in Eq.~\eqref{eq:tildeqgate}, we have $\tilde{u}= (v_4^T\otimes v_2)\tilde{u}_0(v_3\otimes v_1^T)$, where
\begin{equation}
    \tilde{u}_0=\frac{1}{D}\sum_{0\leq p,q\leq D-1} \theta_{p,q}\tau_{p,q}\otimes\tau^*_{p,q},
\end{equation}
with $\tau_{p,q}\equiv\tau^p\sigma^q$. Then the unitarity condition~\eqref{eq:algebric_dual_unitary} on $\tilde{u}$ is equivalent to 
\begin{equation}\label{eq:simplifyunitarityClifford}
    \sum_{0\leq p,q,r,s\leq D-1} \theta^*_{p,q}\theta_{r,s}\tau^\dagger_{p,q}\tau_{r,s}\otimes\tau^T_{p,q}\tau^*_{r,s}=D^2 \tau_{0,0}\otimes\tau_{0,0}.
\end{equation}
Notice that the single site unitary gates $v_{1},v_{2},v_{3},v_{4}$ do not appear in the above expression. We simplify Eq.~\eqref{eq:simplifyunitarityClifford} further with the following relations satisfied by $\tau_{p,q}$
\begin{eqnarray}\label{eq:relations_tau_pq}
    \tau_{p,q}\tau_{r,s}&=&\omega^{qr}\tau_{p+r,q+s},\nonumber\\
    \tau^*_{p,q}&=&\tau_{p,-q},\nonumber\\
    \tau^T_{p,q}&=&\omega^{-pq}\tau_{-p,q}.
\end{eqnarray}
They follow from Eqs.~\eqref{eq:ZnCliffordCR} and~\eqref{eq:def_sigmatau} by straightforward computation. Simplifying Eq.~\eqref{eq:simplifyunitarityClifford} using Eq.~\eqref{eq:relations_tau_pq}, and comparing the coefficients of both sides using the fact that $\{\tau_{p,q}\}_{0\leq p,q\leq D-1}$ forms a basis of the matrix algebra $M_D(\mathbb{C})$, we obtain 
\begin{equation}\label{eq:DUcondition_theta}
    \sum_{0\leq p,q\leq D-1} \theta_{p,q}^* \theta_{p+k,q+l} =0,\text{ for }(k,l)\neq (0,0).
\end{equation}
In this way, the original dual unitarity condition, which involves $2D^4$ equations and $2D^4-1$ real unknowns simplifies to a set of $D^2-1$ equations with $D^2-1$ real unknowns~(notice that we can set $\theta_{0,0}=1$ without loss of generality).  A simple yet nontrivial ansatz for $\theta_{p,q}$ is~\footnote{A particularly simple solution to Eq.~\eqref{eq:DUcondition_theta} is
$\theta_{p,q}=\theta_{p+q}\omega^{p^2+pq}$,
where $\{\theta_{p}\}_{p=0}^{D-1}$ are arbitrary U$(1)$ phases. However, this family of dual-unitary gates are actually the same as those given in Eq.~(25) of Ref.~\cite{marton2022construction}.}
\begin{equation}\label{eq:omega_quadratic}
    \theta_{p,q}=\omega^{\lambda p^2+\mu pq+\nu q^2},
\end{equation}
where $\mu\in\mathbb{Z}$, and $\lambda,\nu\in\mathbb{Z}$ if $D$ is odd while $\lambda,\nu\in\mathbb{Z}/2$ if $D$ is even~(which guarantees that $\theta_{p,q}$ is periodic both in $p$ and $q$ with period $D$). This ansatz also results in the perfect tensors in odd dimensions which are found in \citep{aravinda2021from}. Inserting the ansatz~\eqref{eq:omega_quadratic} into Eq.~\eqref{eq:DUcondition_theta}, we see that dual-unitarity requires that $k=l=0$ is the only solution to the following system of equations~\footnote{A sufficient condition for this is that the determinant $4\lambda\nu-\mu^2$ is invertible modulo $D$. }
\begin{eqnarray}
    2\lambda k+\mu l&=&0~(\mathrm{mod}~D),\nonumber\\
    \mu k+2\nu l&=&0~(\mathrm{mod}~D).
\end{eqnarray}
For example, when $D=3$, $\lambda=\mu=1,\nu=-1$ satisfies this condition. In later sections we will use the ansatz Eq.~\eqref{eq:CliffordParametrization} and Eq.~\eqref{eq:omega_quadratic} to find examples of hierarchical generalizations of dual-unitary gates. 

In this subsection we recapped the basics of dual-unitarity and introduced a novel family of dual-unitary models for $D>2$. A particular subset of solutions from this family appeared before in~\cite{marton2022construction}. This leaves us in a good position to introduce the generalization in the next subsection.

\subsection{Hierarchical Generalization}\label{sec:HG}

As mentioned in the Introduction, dual unitarity imposes conditions on only a single gate, which restricts the possible physical behaviours. It also excludes certain fundamental and well-known gates, such as the Identity and the Controlled-Not gates, which are solvable yet not dual-unitary. 
To unveil more intricate quantum dynamics and include these Clifford gates in a more general notion of solvability,
 we define a \emph{hierarchy of conditions}. This gives us new families of models.
 

Since only one green box plays a part in the dual-unitary condition
(\ref{eq:figure_dual_unitary_condi}), we will call dual-unitarity  also the \emph{first level of the Hierarchy} and denote it as $\mathfrak{L}_1$.
In the subsequent subsections \ref{subsec:Second-Hierarchy}
and \ref{subsec:Third-Hierarchy}, we extend the concept of dual unitarity
($\mathfrak{L}_1$) to conditions involving two and three gates, resulting in the second level $\mathfrak{L}_2$ and
the third level $\mathfrak{L}_3$ of the Hierarchy. 

\subsection{Second level of the Hierarchy\label{subsec:Second-Hierarchy}}

In this subsection, we introduce the gates from $\mathfrak{L}_2$, which are more general than dual-unitary gates. $\mathfrak{L}_2$ contains CNOT and identity, as well as a large family of non-trivial gates whose dynamics cannot be solved by any previous techniques and reveals richer physics. The gates from this family fulfill a condition involving two gates, which is weaker than a dual-unitary condition. 
Here we focus on the case when this condition holds both from the left and the right, but we comment on the non-symmetric case in Sec.~\ref{sec:higher}. The two conditions are:
\begin{equation}
\begin{tikzpicture}[baseline=(current  bounding  box.center), scale=0.7]
\Wgategreen{-0.5}{0.5}
\Wgategreen{0.5}{-0.5}
\MYcircle{0}{1}
\MYcircle{1}{0}
\MYcircle{1}{-1}
\end{tikzpicture}
=
\begin{tikzpicture}[baseline=(current  bounding  box.center), scale=0.7]
\draw[thick] (0,-1)--(1,-1);
\Wgategreen{-0.5}{0.5}
\MYcircle{0}{1}
\MYcircle{0}{00}
\MYcircle{1}{-1}
\end{tikzpicture}
,
\qquad
\begin{tikzpicture}[baseline=(current  bounding  box.center), scale=0.7]
\Wgategreen{0.5}{0.5}
\Wgategreen{-0.5}{-0.5}
\MYcircle{0}{1}
\MYcircle{-1}{0}
\MYcircle{-1}{-1}
\end{tikzpicture}
=
\begin{tikzpicture}[baseline=(current  bounding  box.center), scale=0.7]
\draw[thick] (0,-1)--(-1,-1);
\Wgategreen{0.5}{0.5}
\MYcircle{0}{1}
\MYcircle{0}{0}
\MYcircle{-1}{-1}
\end{tikzpicture}
.\label{eq2:bottomtotop}
\end{equation}
Algebraically, we can express the condition as 
\begin{equation}
\begin{aligned}(I_{D}\otimes\tilde{u}^{\dagger})\cdot\tilde{u}^{\dagger}\tilde{u}\otimes I_{D}\cdot(I_{D}\otimes\tilde{u})=I_{D}\otimes\tilde{u}^{\dagger}\tilde{u},\\
(I_{D}\otimes\tilde{u})\cdot\tilde{u}\tilde{u}^{\dagger}\otimes I_{D}\cdot(I_{D}\otimes\tilde{u}^{\dagger})=I_{D}\otimes\tilde{u}\tilde{u}^{\dagger}.
\end{aligned}
\label{eq:2ndEq}
\end{equation}
A direct observation shows that if a circuit is $\mathfrak{L}_1$, it must be $\mathfrak{L}_2$. This, together with the fact that the identity is in $\mathfrak{L}_2$ but not in $\mathfrak{L}_1$, implies that $\mathfrak{L}_1$ is a proper subset of $\mathfrak{L}_2 \ $:
$\mathfrak{L}_1\subsetneqq\mathfrak{L}_2$. In the following we focus on the gates that are in $\mathfrak{L}_2$ but not in $\mathfrak{L}_1$, the set we denote as $\overline{\mathfrak{L}}_2=\mathfrak{L}_2-\mathfrak{L}_1$. 
 Note that a necessary condition for a gate $u$ to be in $\overline{\mathfrak{L}}_2$ is that  $\tilde{u}$ is not invertible, since otherwise one can contract both sides of Eq.~\eqref{eq2:bottomtotop} with $\tilde{u}^{-1}$ and recover the dual unitary condition. 

Similarly to the $\mathfrak{L}_1$ case~\cite{bertini2019exact},
we can figure out the complete parametrization of $\overline{\mathfrak{L}}_2$ for qubits. As explained in App. \ref{app:qubitsDetails}, we used analytical analysis with
the numerical help of Mathematica.
When $D=2$,
an exhaustive parametrization of 2-qubit gates is
\begin{equation}
u=v_{1}\otimes v_{2} \ e^{i(J_{x}\sigma_{x}\sigma_{x}+J_{y}\sigma_{y}\sigma_{y}+J_{z}\sigma_{z}\sigma_{z})} \ v_{3}\otimes v_{4}.
\label{eq:parameter2qubit}\end{equation}
Here $\sigma_j$ are Pauli matrices and $v_{1},v_{2},v_{3},v_{4}$ are all single site
gates from $\mathbb{SU}(2)$, and $0\leq J_j < \pi/2$. 
We may simplify the gate structure by setting $v_{3}=v_{4}=I_{D}$
without any loss of generality~\footnote{This is true because because $v_{1}$ at this time step can be
combined with $v_{4}$ from the next time step, allowing for the redefinition
$v_{1}\to v_{4}\cdot v_{1}$. This reasoning also applies to $v_{2}$ and $v_{3}$.}.
The trivial example from $\overline{\mathfrak{L}}_2$ is a tensor product of two single-site operators.
Apart from that, the $\overline{\mathfrak{L}}_2$ 
condition fixes $J_{z}=\frac{\pi}{4},J_{x}=J_{y}=0$~\footnote{The permutations among $x,y,z$ also work. 
} and $v_{1},v_{2}$ to be elements of the set 
\begin{equation}
\{U(r,\theta,\phi)|\sqrt{2}\sin r\sin\theta=\pm1\}, 
\label{eq:condition2ndqubit}
\end{equation}
where $U(r,\theta,\phi)$ is defined as $e^{ir(\cos\theta \ \sigma_{z}+\sin\theta\ \cos\phi \ \sigma_{x}+\sin\theta \ \sin\phi \ \sigma_{y})}$, 
representing a $\mathbb{SU}(2)$ on the Bloch sphere. Geometrically,
this specific combination of $r,\theta,\phi$ represents a rotation that
maps $\sigma_{z}$ to the $x-y$ plane.

The dimension of $\overline{\mathfrak{L}}_2$ can be counted as follows. Out of $12$ parameters determining the $4$ local $\mathbb{SU}(2)$ gates, e.g. the Euler angles, two are redundant because the rotation around the $z-$axis commutes with the Ising interaction resulting from $J_z=\frac{\pi}{4},J_x=J_y=0$. Further, Eq. (\ref{eq:condition2ndqubit}) provides $2$ constraints. After considering the global phase, the total independent parameters to characterize a qubit $\overline{\mathfrak{L}}_2$ circuit is $12-2-2+1$. Therefore, we have defined a new $9$-dimensional 
family of solvable models which are not part of $12$-dimensional set of $\mathfrak{L}_1$ gates~\cite{prosen2021manybody}.

Note that the control not gate ($\mathrm{CNOT}$) can be decomposed into the form of Eq. (\ref{eq:parameter2qubit}) (with $v_4$ different than identity) as:
\begin{equation}
\begin{aligned}
v_1 & =e^{-i\frac{\pi}{4}\sigma_{z}},\; & v_2 & = H\sigma_{x}\cdot e^{i\frac{\pi}{4}\sigma_{z}},\\
v_3 & =I_2,\; & v_4 & =\sigma_{x}H,
\end{aligned}
\end{equation}
with $J_x=0,J_y=0,J_z=\frac{\pi}{4}$ and an addition global phase $e^{-i\frac{\pi}{4}}$.
$H=\frac{1}{\sqrt{2}}\begin{pmatrix}1 & 1\\
1 & -1
\end{pmatrix}$ is the Hadamard gate. To check that this satisfy~\eqref{eq:condition2ndqubit}, we include $v_{4}$ back in the previous layer of the gates therefore obtaining combined(c) $(v_{4})_c = I_2$, $(v_1)_c = v_4 v_1$. 

In the case of bigger local dimensions $D$, we do not yet possess a complete parametrization
for the $\mathfrak{L}_2$ case. Nevertheless, we can discern two
distinct and rich families. The first family is associated with generalized
Controlled-NOT gate in higher dimension surrounded by $4$ single site operators
\begin{equation}
u=v_1 \otimes v_2 \ \ C_{\tau} \ \ v_3 \otimes v_4,
\end{equation}
with 
$C_{\tau}=\sum_{i}\ket{i}\bra{i}\otimes \tau^{i}$.
Following a similar argument as below Eq. (\ref{eq:parameter2qubit}), we set $v_3=v_4=I_D$. In this case, $v_{1}$ and $v_{2}$
must satisfy
\begin{equation}
\begin{aligned}\sum_{j}\bra{j}v_{1}\ket{i}\bra{j+k'-k}v_{1}^{*}\ket{i} & =\delta_{k,k'}\ \mathrm{for}\ \forall k,k',i,\\
\sum_{j}\bra{i}v_{2}\ket{j}\bra{i}v_{2}^{*}\ket{j+k'-k} & =\delta_{k,k'}\ \mathrm{for}\ \forall k,k',i.
\end{aligned}
\end{equation}
These two equations share a symmetry of exchanging columns and rows between themselves.

The second family is derived using the Clifford group method from subsection~\ref{sec:DU}. Utilizing the proposed ansatz from Eqs. (\ref{eq:2quditUgeneralform}) and (\ref{eq:CliffordParametrization}), we set $v_3=v_4=I_D$ and simplify Eq. (\ref{eq:2ndEq}) to: 
\begin{widetext}
\begin{equation}
\begin{aligned}\left(\sum_{b}\theta_{p_{b},q_{b}}^{*}\theta_{p_{b}+k,q_{b}+l}\right)\left(\sum_{d}\theta_{p_{d},q_{d}}^{*}\theta_{p_{d}+s,q_{d}+t}\tau_{p_{d},q_{d}}^{\dagger}v_{1}^{*}\tau_{k,-l}v_{1}^{T}\tau_{p_{d},q_{d}}\right)=0,\ \\
\left(\sum_{b}\theta_{p_{b},q_{b}}^{*}\theta_{p_{b}+k,q_{b}+l}\right)\left(\sum_{d}\theta_{p_{d},q_{d}}^{*}\theta_{p_{d}+s,q_{d}+t}\tau_{-p_{d},q_{d}}^{\dagger}v_{2}^{*}\tau_{-k,-l}v_{2}^{T}\tau_{-p_{d},q_{d}}\right)=0.
\end{aligned}\label{eq:Clifford_parametrization_2nd_result}
\end{equation}
\end{widetext}
Here $\sum_b$ is a shorthand for $\sum_{0\leq p_b,q_b\leq D-1}$. 
The above equation should hold for $\forall(s,t)\neq(0,0)\ \mathrm{and}\ (k,l)\neq(0,0)$. If all terms in the first sum vanish separately, we obtain the $\mathfrak{L}_1$. 
From these nonlinear equations, we can derive a family of $\overline{\mathfrak{L}}_2$, which is just one of the many possible solutions.
The family is defined for $ D=4k+2$ as
\begin{equation}
 k\in\mathbb{N}^{+}\, \; \mathrm{ and }\,\; \theta_{p,q}=\omega^{\frac{Dpq}{2}} \, .
    \label{eq:simplest_2nd_Hierarchy_case}
\end{equation}
Another nontrivial example is given by 
\begin{equation}
\theta_{p,q}=\begin{cases}
\omega^{\frac{p^{2}}{2}}, & D=\mathrm{even},\\
\omega^{p^{2}}, & D=\mathrm{odd}.
\end{cases}\; \label{eq:second_example}\end{equation}
In both of the two examples, the so far unspecified $v_1$ and $v_2$ belong to a non-trivial subset of $\mathbb{SU}(D)$. Finding all possible $v_1$ and $v_2$ is, in general, hard. One can try guessing good candidates and check if the conditions in Eq.~(\ref{eq:Clifford_parametrization_2nd_result}) are satisfied. There is a way of simplifying the conditions using the structure of the Clifford ansatz, which help in deducing  $v_1$ and $v_2$, see App. \ref{app:quditsDetails} for the details. 
Different choices lead to both ergodic and non-ergodic dynamics.

Before concluding this subsection, we would like to highlight that the two equalities in Eq. (\ref{eq2:bottomtotop}) and unitarity
additionally imply
\begin{equation}
\begin{tikzpicture}[baseline=(current  bounding  box.center), scale=0.7]
\Wgategreen{-0.5}{0.5}
\Wgategreen{0.5}{-0.5}
\MYcircle{-1}{1}
\MYcircle{-1}{0}
\MYcircle{0}{-1}
\end{tikzpicture}
=
\begin{tikzpicture}[baseline=(current  bounding  box.center), scale=0.7]
\Wgategreen{0.5}{-0.5}
\draw[thick] (0,1)--(-1,1);
\MYcircle{-1}{1}
\MYcircle{0}{0}
\MYcircle{0}{-1}
\end{tikzpicture}
,
\qquad
\begin{tikzpicture}[baseline=(current  bounding  box.center), scale=0.7]
\Wgategreen{-0.5}{-0.5}
\Wgategreen{0.5}{0.5}
\MYcircle{1}{1}
\MYcircle{1}{0}
\MYcircle{0}{-1}
\end{tikzpicture}
=
\begin{tikzpicture}[baseline=(current  bounding  box.center), scale=0.7]
\Wgategreen{-0.5}{-0.5}
\draw[thick] (0,1)--(1,1);
\MYcircle{1}{1}
\MYcircle{0}{0}
\MYcircle{0}{-1}
\end{tikzpicture}
,
\label{eq:2ndfig}
\end{equation}
separately. i.e., a Hierarchical condition along with unitarity implies its $180^\circ$ rotation version.
The proof is shown in App. \ref{app:proof2nd}. 
 Eq. (\ref{eq:2ndfig}) will play an important role in computing the spatio-temporary correlation functions.

\subsection{Third level of the Hierarchy\label{subsec:Third-Hierarchy}}

Following the principles from subsection \ref{subsec:Second-Hierarchy}, we define the third level hierarchical condition for $\mathfrak{L}_3$ as 
\begin{equation}
\begin{tikzpicture}[baseline=(current  bounding  box.center), scale=0.7]
\Wgategreen{-1.5}{0.5}
\Wgategreen{-0.5}{-0.5}
\Wgategreen{-2.5}{1.5}
\MYcircle{-1}{1}
\MYcircle{0}{0}
\MYcircle{0}{-1}
\MYcircle{-2}{2}
\end{tikzpicture}
=
\begin{tikzpicture}[baseline=(current  bounding  box.center), scale=0.7]
\Wgategreen{-1.5}{0.5}
\Wgategreen{-2.5}{1.5}
\draw[thick] (-1,-1)--(0,-1);
\MYcircle{0}{-1}
\MYcircle{-1}{0}
\MYcircle{-1}{1}
\MYcircle{-2}{2}
\end{tikzpicture}
.\label{eq:3rdHierarchydefinition}
\end{equation}An immediate observation reveals that a gate characterized as $\mathfrak{L}_2$ is also the $\mathfrak{L}_3$. Nonetheless, we are again interested
in the special subset of $\mathfrak{L}_3$ which does not belong to
$\mathfrak{L}_2$, designated as $\overline{\mathfrak{L}}_3=\mathfrak{L}_3-\mathfrak{L}_2$.
A notable example within $\overline{\mathfrak{L}}_3$ is the controlled-Z gate.

We again use the complete parameterization of 2-qubit gates (\ref{eq:parameter2qubit}) and wlog set $v_3=v_4=I_2$. The condition defining $\overline{\mathfrak{L}}_3$ is satisfied either for all diagonal gates or for the case where $J_{x}=J_{y}=0$ and any $J_z$ with $v_i$ satisfying $v_i=\cos{\phi_i}\sigma_x+\sin{\phi_i}\sigma_y,i\in\{1,2\}$. 

To get some examples for $D>2$, we use our Clifford gate parametrization method
from Secion~\ref{sec:DU}. 
The algebraic equation
of $\theta_{p,q}$ can be found in the App.~\ref{sec:appendixA}. To obtain some examples, we take the single-site operators $v_i$ to be the identity. Some classes of the solutions obtained in this way are shown below.
\begin{equation}
\theta_{p,q}=\begin{cases}
\omega^{p^{2}+\frac{3}{2}q^{2}}, & D=12m+2,\\
\omega^{p^{2}+q^{2}}, & D=8m+4,\\
\omega^{p^{2}+\frac{3}{2}q^{2}}, & D=12m+6,m\neq1\ \mathrm{mod}\ 3,\\
\omega^{p^{2}+\frac{3}{2}q^{2}}, & D=12m+10.
\end{cases}
\end{equation}

In principle, nothing stops us from going beyond the $\mathfrak{L}_3$, by demanding even more general condition with even more gates. We expect the examples to be constructed in a similar way.
\section{Applications}
\label{sec:applications}
\subsection{Spatio-temporary correlator functions}\label{subsec:STCF}
In this subsection we focus on the spatio-temporal correlation functions, which are the most common objects to characterize the dynamics. In particular, they provide information about the thermalization and ergodicity of the system. 

In most cases, the exact non-perturbative calculation of the spatio-temporary correlators is only available in free models and to some extent in interacting integrable ones~\cite{Medenjak_2017,Klobas_2019,Klobas_2021}. 
Important progress has been made in understanding the correlations also in chaotic models, in particular dual-unitary ($\mathfrak{L}_1$) circuits which we extend here.

Due to the trivial propagation of an identity operator, we are only interested in the correlation function between two traceless Hilbert-Schmidt normalized operators $a_{i},b_{j}$\footnote{Hilbert Schmidt normalized means that $\mathrm{Tr}a_i^\dagger a_i=1$}.
Working in the Heisenberg picture, the spatio-temporal correlation function of normalized local operators can be expressed as 
\begin{equation}
C_{ij}(t)=\langle a_{i}(t)b_{j}\rangle=D\mathrm{Tr}\left((\mathbb{U}^t)^{\dagger} a_{i}\mathbb{U}^tb_{j}\frac{1}{D^{2L}}\right).
\label{eq:definitionofcorrelationfunction}
\end{equation}
The factor $\frac{1}{D^{2L}}$ comes from the normalized infinite temperature state $\rho_{\infty}=\frac{I_{D^{2L}}}{D^{2L}}$. We also include a prefactor $D$ in the definition to ensure that the autocorrelation function at time $0$ is normalized to $1$. Alternatively, this can be viewed as a quench from the $b_j \1$ state, i.e. $b_j$ applied to the maximally mixed state. 
The correlations in the folded picture are graphically
expressed as 
\begin{align}
&C_{ij}(t)=
\begin{tikzpicture}[baseline=(current bounding box.center), scale=0.55]
\foreach \jj[evaluate=\jj as \j using -2*(ceil(\jj/2)-\jj/2)] in {0,...,5}
\foreach \i in {1,...,6}
{\Wgategreen{2*\i+\j}{\jj}}
\foreach \i in {2,...,13}{
\MYcircle{\i-.5}{-0.5}
\MYcircle{\i-1.5}{6-0.5}
}
\MYcircleB{3.5}{-.5}
\MYcircleB{5.5}{6-.5}
\Text[x=3.5,y=-1.0]{$b$}
\Text[x=5.5,y=6.0]{$a$}
\end{tikzpicture}\, . \label{eq:Corr1}
\end{align}
Employing the time unitarity enables us to simplify the circuit from the bottom
and top, yielding

\begin{align}
&C_{ij}(t)=
\begin{tikzpicture}[baseline=(current bounding box.center), scale=0.55]
\Wgategreen{3}{0}
\Wgategreen{2}{1}
\Wgategreen{4}{1}
\Wgategreen{1}{2}
\Wgategreen{3}{2}
\Wgategreen{5}{2} 
\Wgategreen{2}{3}
\Wgategreen{4}{3}
\Wgategreen{6}{3}
\Wgategreen{3}{4} 
\Wgategreen{5}{4}
\Wgategreen{4}{5}
\MYcircle{1.5}{0.5}
\MYcircle{0.5}{1.5}
\foreach \i in {1,...,4}
{\MYcircle{2.5+\i}{\i-1.5}}
\foreach \i in {1,...,4}
{
\MYcircle{4.5-\i}{6.5-\i}
}
\MYcircle{5.5}{4.5}
\MYcircle{6.5}{3.5}
\MYcircleB{2.5}{-.5}
\MYcircleB{4.5}{6-.5}
\Text[x=2.5,y=-1.0]{$b$}
\Text[x=4.5,y=6.0]{$a$}
\end{tikzpicture}\, . 
\label{eq:2pointaftertimeuni}
\end{align}

\subsubsection{Dual unitarity}
For completeness, here we briefly summarize the result for dual-unitary circuit from~\cite{bertini2019exact}. 
Intuitively, the time unitarity and space unitarity both determine a light cone outside which the correlation function vanishes. Therefore, correlators can solely manifest at the intersection of these two cones, forming a 1-dimensional straight line that precisely bisects the temporal and spatial directions
\begin{equation}\label{eq:diagonal}
\begin{aligned}
&C_{ij}(t)=
\begin{tikzpicture}[baseline=(current bounding box.center), scale=0.55]
\foreach \i in {0,...,5}
{\Wgategreen{3+\i}{\i}}
\foreach \i in {0,...,5}
{\MYcircle{2.5+\i}{0.5+\i}}
\foreach \i in {0,...,5}
{\MYcircle{3.5+\i}{-0.5+\i}}
\MYcircleB{2.5}{-.5}
\MYcircleB{8.5}{6-.5}
\Text[x=2.5,y=-1.0]{$b$}
\Text[x=8.5,y=6.0]{$a$}
\end{tikzpicture}
,
\end{aligned}
\end{equation}
with other correlators vanishing. That other correlations vanish can be seen by repeatedly applying~\eqref{eq:figure_dual_unitary_condi} to expression~\eqref{eq:2pointaftertimeuni}, which results in $\braket{\mcirc|\mcircf}=0$ since the operators $a$ and $b$ are traceless.
In Eq.~\eqref{eq:diagonal} each time step is just a quantum channel over $D\times D$ Hilbert space. Therefore, the correlators for the $\mathfrak{L}_1$ circuits can always be calculated efficiently and propagates only along two directions with maximal speed.

\subsubsection{$\overline{\mathfrak{L}}_2$ circuits}
Moving beyond the $\mathfrak{L}_1$, we are interested in which new features appear in the $\overline{\mathfrak{L}}_2$ circuits.
Said differently, we are interested in what happens if the weaker condition~\eqref{eq2:bottomtotop} defining $\mathfrak{L}_2$ is satisfied but dual unitarity condition~\eqref{eq:figure_dual_unitary_condi} is not.

We apply Eq. (\ref{eq2:bottomtotop}) to Eq. (\ref{eq:2pointaftertimeuni}) and further simplify the circuit to
\begin{align}
&C_{ij}(t)=
\begin{tikzpicture}[baseline=(current bounding box.center), scale=0.55]
\Wgategreen{4}{5}
\Wgategreen{5}{4}
\Wgategreen{4}{3}
\foreach \i in {1,2,3}
{\Wgategreen{2+\i}{\i-1}}
\foreach \i in {1,...,5}
{
\MYcircle{3.5}{\i+0.5}
}
\foreach \i in {1,2,3}
{\MYcircle{5.5}{1.5+\i}}
\MYcircle{2.5}{0.5}
\foreach \i in {1,2,3}
{\MYcircle{2.5+\i}{\i-1.5}}
\MYcircleB{2.5}{-.5}
\MYcircleB{4.5}{6-.5}
\Text[x=2.5,y=-1.0]{$b$}
\Text[x=4.5,y=6.0]{$a$}
\end{tikzpicture}\, .
\end{align}
Lastly, Eq. (\ref{eq:2ndfig}) is utilised to address the corner of the
path, leading to\begin{align}
&C_{ij}(t)=
\begin{tikzpicture}[baseline=(current bounding box.center), scale=0.55]
\Wgategreen{4}{5}
\Wgategreen{5}{4}
\Wgategreen{4}{3}
\foreach \i in {1,2}
{\Wgategreen{2+\i}{\i-1}}
\foreach \i in {1,...,5}
{
\MYcircle{3.5}{\i+0.5}
}
\foreach \i in {2,3}
{\MYcircle{5.5}{1.5+\i}}
\MYcircle{2.5}{0.5}
\foreach \i in {1}
{\MYcircle{2.5+\i}{\i-1.5}}
\foreach \i in {1,2,3}
{\MYcircle{4.5}{\i-0.5}}
\MYcircleB{2.5}{-.5}
\MYcircleB{4.5}{6-.5}
\Text[x=2.5,y=-1.0]{$b$}
\Text[x=4.5,y=6.0]{$a$}
\end{tikzpicture}\, .
\end{align}
This correlator vanishes because the discontinuous path will be simplified to $\mathrm{\mathrm{Tr}}a_{i}\mathrm{Tr}b_{j}$ and both are traceless according to our assumption.

Therefore, the existence of nonvanishing correlators is limited to three possible directions, either at the light cone or at velocity zero. 

\begin{equation}
\begin{aligned}
&C_{ij}(t)=
\begin{tikzpicture}[baseline=(current bounding box.center), scale=0.55]
\foreach \i in {0,...,5}
{\Wgategreen{3+\i}{\i}}
\foreach \i in {0,...,5}
{\MYcircle{2.5+\i}{0.5+\i}}
\foreach \i in {0,...,5}
{\MYcircle{3.5+\i}{-0.5+\i}}
\MYcircleB{2.5}{-.5}
\MYcircleB{8.5}{6-.5}
\Text[x=2.5,y=-1.0]{$b$}
\Text[x=8.5,y=6.0]{$a$}
\end{tikzpicture}
,\\
&C_{ij}(t)=
\begin{tikzpicture}[baseline=(current bounding box.center), scale=0.55]
\foreach \i in {0,1,2}
{
\Wgategreen{4}{2*\i}
\Wgategreen{3}{1+2*\i}
\MYcircle{4.5}{2*\i-0.5}
\MYcircle{4.5}{2*\i+0.5}
\MYcircle{2.5}{1.5+2*\i}
\MYcircle{2.5}{0.5+2*\i}
}
\MYcircleB{3.5}{-.5}
\MYcircleB{3.5}{6-.5}
\Text[x=3.5,y=-1.0]{$b$}
\Text[x=3.5,y=6.0]{$a$}
\end{tikzpicture}.
\end{aligned}
\label{eq:finalresult1site}
\end{equation}
These expressions can be written using four single qudit channels: 
\begin{equation}
\begin{aligned}
&\epsilon_L(b)=
\begin{tikzpicture}[baseline=(current bounding box.center), scale=0.55]
\Wgategreen{0}{0}
\MYcircle{-0.5}{-0.5}
\MYcircle{-0.5}{0.5}
\MYcircleB{0.5}{-0.5}
\Text[x=0.5,y=-1]{$b$}
\end{tikzpicture}
,
\qquad
&\epsilon_R(b)=
\begin{tikzpicture}[baseline=(current bounding box.center), scale=0.55]
\Wgategreen{0}{0}
\MYcircle{0.5}{-0.5}
\MYcircle{0.5}{0.5}
\MYcircleB{-0.5}{-0.5}
\Text[x=-0.5,y=-1]{$b$}
\end{tikzpicture}
,\\
&M_L(b)=
\begin{tikzpicture}[baseline=(current bounding box.center), scale=0.55]
\Wgategreen{0}{0}
\MYcircle{0.5}{-0.5}
\MYcircle{-0.5}{0.5}
\MYcircleB{-0.5}{-0.5}
\Text[x=-0.5,y=-1]{$b$}
\end{tikzpicture}
,
\qquad
&M_R(b)=
\begin{tikzpicture}[baseline=(current bounding box.center), scale=0.55]
\Wgategreen{0}{0}
\MYcircle{-0.5}{-0.5}
\MYcircle{0.5}{0.5}
\MYcircleB{0.5}{-0.5}
\Text[x=0.5,y=-1]{$b$}
\end{tikzpicture}
.
\end{aligned}
\end{equation}
To simplify the analysis, we assume $j$ to be an integer, and the other case follows analogously. Thus
\begin{equation}
C_{ij}(t)\!\!=\!\!\begin{cases}
\!\mathrm{Tr}\left(aM_{L}^{2t}(b)\right), & \!\!\!\!\!t=i-j,\\
\!\mathrm{Tr}\left(a(\epsilon_{R})^{k}(\epsilon_{L}\epsilon_{R})^{\lfloor\frac{t}{2}\rfloor}(b)\right), &
\!\!\!\!\!i\!=\!j,t\!=\!\mathbb{Z}\!+\!\frac{k}{2},\\
\!0, & \!\!\!\!\!\mathrm{otherwise}.
\end{cases}\label{eq:expression_CF_1site}
\end{equation}
This $C_{ij}(t)$ behaves differently than
that in the dual-unitary case~\cite{bertini2019exact}, 
as the circuits from $\mathfrak{L}_2$ allow for an additional non-vanishing direction along the time axis. 

Let us mention here the connection with tri-unitaries circuits proposed in \cite{jonay2021triunitary}, where the correlation function also exclusively manifest in the same three directions. In fact, we can group the legs of two 2-qubit gates into a 3-qubit gates as
\begin{equation}
\begin{tikzpicture}[baseline=(current bounding box.center), scale=0.55]
    \Wgategreen{0}{0}
    \Wgategreen{-1}{-1}
    \draw[red!70!magenta,very thick] (-1.25,-1.25)--(-1.5,-1.5);
    \draw[red!70!magenta,very thick] (-0.75,-1.25)--(-0.5,-1.5);
    \draw[red!70!magenta,very thick] (0.25,-0.25)--(0.5,-0.5);
    \draw[skyblue, very thick] (-1.25,-0.75) -- (-1.5,-0.5);
    \draw[skyblue, very thick](-0.25,0.25) -- (-0.5,0.5);
    \draw[skyblue, very thick](0.25,0.25) -- (0.5,0.5);
\end{tikzpicture}
\Rightarrow
\begin{tikzpicture}[baseline=(current bounding box.center), scale=0.55]
 \fill[mygreen] (0,0) -- (0.35,0.35) -- (0.35,0.85) -- (0,1.2) -- (-0.35,0.85) -- (-0.35,0.35) -- cycle;
 \draw[red!70!magenta,very thick](0,0)--(0,-0.25);
 \draw[red!70!magenta,very thick](0.35,0.35)--(0.6,0.1);
 \draw[red!70!magenta,very thick](-0.35,0.35)--(-0.6,0.1);
 \draw[skyblue, very thick](0,1.2)--(0,1.45);
 \draw[skyblue, very thick](0.35,0.85) -- (0.6,1.1);
 \draw[skyblue, very thick](-0.35,0.85) -- (-0.6,1.1);
\end{tikzpicture}\; .
\end{equation}
However, in the tri-unitary case, the condition is
\begin{equation}
\begin{tikzpicture}[baseline=(current bounding box.center), scale=0.55]
 \fill[mygreen] (0,0) -- (0.35,0.35) -- (0.35,0.85) -- (0,1.2) -- (-0.35,0.85) -- (-0.35,0.35) -- cycle;
 \draw[very thick](0,0)--(0,-0.25);
 \draw[very thick](0.35,0.35)--(0.6,0.1);
 \draw[very thick](-0.35,0.35)--(-0.6,0.1);
 \draw[very thick](0,1.2)--(0,1.45);
 \draw[very thick](0.35,0.85) -- (0.6,1.1);
 \draw[very thick](-0.35,0.85) -- (-0.6,1.1);
 \MYcircle{0.6}{0.1}
 \MYcircle{0}{-0.25}
 \MYcircle{0.6}{1.1}
\end{tikzpicture}
=
\begin{tikzpicture}[baseline=(current bounding box.center), scale=0.55]

\draw[very thick](-0.5,1)--(0.5,1);
\draw[very thick](-0.5,0.5)--(0.5,0.5);
\draw[very thick](-0.5,0)--(0.5,0);
\MYcircle{0.5}{1}
\MYcircle{0.5}{0.5}
\MYcircle{0.5}{0}
\end{tikzpicture},
\end{equation}
which is a much stronger condition than Eq. (\ref{eq2:bottomtotop}).

Interestingly, in the context of qubits ($D=2$), it is impossible to observe all in principle allowed physical behaviors.
The correlations along the light rays vanish since channels $\epsilon_{L}$ and $\epsilon_{R}$ correspond to the total depolarizing channel.
Nevertheless, when $D>2$, there are examples manifesting all of the properties discussed above, i.e. nonvanishing correlations in all three directions at all times. 
In other words, both the correlations in Eq. (\ref{eq:finalresult1site}) are nontrivial. A such example is given in Eq. (\ref{eq:simplest_2nd_Hierarchy_case}) which is also shown in Fig. \ref{Correlation_func_sup2}(a). In this figure, the operator has support on two nearest neighbor sites, to eliminate the odd/even effect (for details see subsec. \ref{subsec:Biggersupports}). 

\subsubsection{$\overline{\mathfrak{L}}_3$ and higher levels}
\label{sec:higher}
\begin{figure}
\includegraphics[width=1.0\columnwidth]{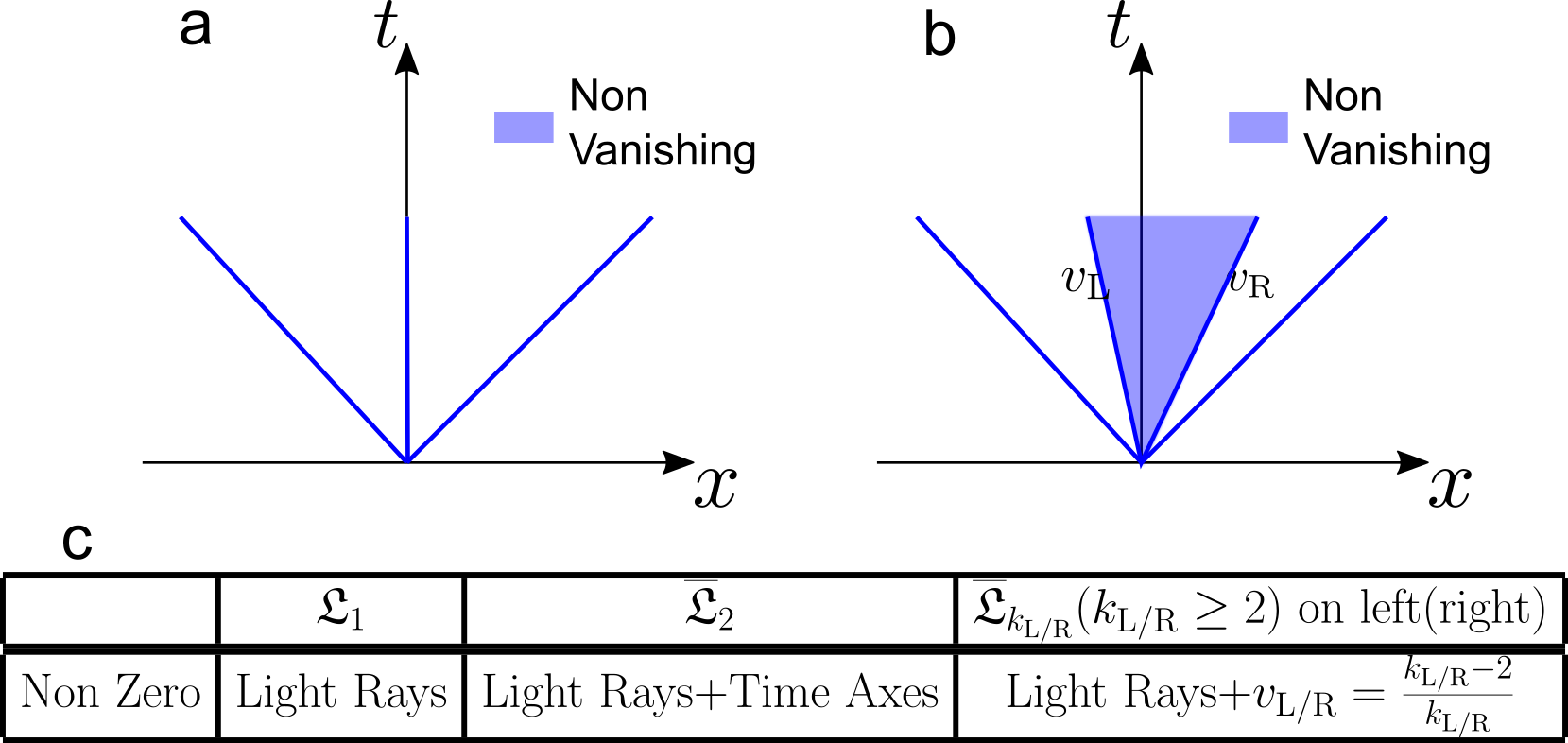}
\caption{
The different Hierarchical conditions lead to different behaviors of 2-point correlation functions. a) $\overline{\mathfrak{L}}_2$ Hierarchical circuits with non-vanishing correlators along three directions. b) General Hierarchical circuits that satisfy conditions for the $k_{\text{L/R}}$ level of Hierarchy from the left and right, respectively.
Their correlators are non-vanishing in the middle part and on the light rays. 
The middle part is restricted to the inner light cone given by the velocity $v$ satisfying $-v_{\text{L}} \leq v \leq v_{\text{R}}$, with $v_{\text{L/R}}$ 
following from $k_{\text{L/R}}$ level Hierarchical conditions. 
c) summary of the property of the two-point correlators. For a circuit without any Hierarchical conditions, we may equivalently consider it as $k=\infty$.}\label{figure1:illustratevelocity}
\end{figure}

In the case where the gate is classified as $\overline{\mathfrak{L}}_3$, the correlation
function is reduced to  
\begin{equation}
C_{ij}(t)=
\begin{tikzpicture}[baseline=(current bounding box.center), scale=0.55]
\foreach \i in {0,...,5}
{\Wgategreen{0}{2*\i}}
\foreach \i in {0,...,5}
{
\Wgategreen{1}{1+2*\i}
}
\foreach \i in {0,...,4}
{\Wgategreen{-1}{1+2*\i}}
\foreach \i in {0,...,3}
{\Wgategreen{2}{4+2*\i}}
\foreach \i in {0,1,2}
{
\Wgategreen{-2}{2+2*\i}
}
\Wgategreen{3}{7}
\foreach \i in {0,1,2}
{
\MYcircle{-1.5}{7.5+\i}
\MYcircle{1.5}{0.5+\i}
\MYcircle{2.5}{3.5+\i}
\MYcircle{2.5}{8.5+\i}
}
\foreach \i in {0,...,5}
{
\MYcircle{-2.5}{1.5+\i}
}
\MYcircle{0.5}{-0.5}
\MYcircle{3.5}{6.5}
\MYcircle{3.5}{7.5}
\MYcircle{1.5}{11.5}
\MYcircle{-0.5}{10.5}
\MYcircleB{-0.5}{-0.5}
\MYcircleB{0.5}{11.5}
\MYcircle{-1.5}{0.5}
\Text[x=-0.5,y=-1] {$b$};
\Text[x=.5,y=12] {$a$};
\end{tikzpicture}
.\label{eq:Non_zero_along_middle_part}
\end{equation}
Intriguingly, this correlator does not vanish within the entire light cone, and no closed expression for it can be derived with a scaling polynomial in system size. Nevertheless, the hierarchical conditions still imply that some of the correlations are zero. As a general rule, for a $k$th level of Hierarchical circuit $\overline{\mathfrak{L}}_k=\mathfrak{L}_k-\mathfrak{L}_{k-1}$, the correlations are nonzero along both the middle part as shown in Eq. (\ref{eq:Non_zero_along_middle_part}) and the maximal velocity light rays. If we focus on the middle part and consider it as the inner light cone, its inner light cone velocity will be suppressed to $\nu_k=\frac{k-2}{k}$ for $k\geq 2$, see Fig. \ref{figure1:illustratevelocity}. Notice that a Hierarchical condition along with unitarity always imply its $180^\circ$ rotation, as explained in Eq. (\ref{eq2:bottomtotop}), Eq. (\ref{eq:2ndfig}) and App. \ref{app:proof2nd}. These pairs of conditions lead to a backward-propagating inner light cone originated from $a$ (the operator at the top) and a forward-propagating inner light cone originated from $b$ (the operator at the bottom), both with velocity $\nu_k$. The 2-point correlator vanishes when these two inner lightcones do not overlap, except when $a$ lies exactly on the light ray of $b$.  

Finally, we also illustrate the possibility of choosing different types of condition (or no condition at all) from the left and right directions, resulting in an asymmetric inner light cone. This possibility is shown and discussed in Fig. \ref{figure1:illustratevelocity}.

\subsection{Bigger operators and higher orders}
\label{subsec:Biggersupports}

In contrast to previous research, which concentrated primarily on correlators supported on a single site, exploring operators with multi-site support sheds light on more intricate underlying physical phenomena. 
Specifically, for correlators supported on multiple sites, their behavior resembles that described in Eq. (\ref{eq:expression_CF_1site}), where correlation functions manifest exclusively along three directions. Remarkably, in the case of qubits, these correlation functions persist over time unlike the single-site supported ones.

Here we present examples of the correlators on nearest neighbor sites. The derivation is essentially the same as in the previous section with some special attention to the simplifications around the operators. The location of a multi-site operator is indexed by its leftmost site. For simplicity, we assume $j$ is an even number and represent the nearest neighbor two-site operator by a black square \begin{tikzpicture}[baseline=(current bounding box.center), scale=0.55]
\MYsquareB{4}{0}
\draw[very thick](3.75,0.25)--(3.5,0.5);
\draw[very thick](4.25,0.25)--(4.5,0.5);
\end{tikzpicture} $=$ \begin{tikzpicture}[baseline=(current bounding box.center), scale=0.55]
\Wgategreen{0}{0}
\fill[pattern=north east lines, pattern color=gray, postaction={draw=black}] (0,-0.5) ellipse (0.75 and 0.1);
\end{tikzpicture}.
\begin{equation}
\braket{a_{j+t-1}(t) b_j}=
\begin{tikzpicture}[baseline=(current bounding box.center), scale=0.55]
\foreach \i in {1,...,5}
{\Wgategreen{3+\i}{\i}}
\foreach \i in {1,...,5}
{\MYcircle{2.5+\i}{0.5+\i}}
\foreach \i in {1,...,5}
{\MYcircle{3.5+\i}{-0.5+\i}}
\MYsquareB{3.0}{0}
\draw[very thick](3,0)--(3.5,0.5);
\draw[very thick](3,0)--(2.5,0.5);
\MYcircle{2.5}{0.5};
\MYsquareB{9}{6}
\draw[very thick](9,6)--(8.5,5.5);
\draw[very thick](9,6)--(9.5,5.5);
\MYcircle{9.5}{5.5};
\Text[x=3,y=-.65]{$b$}
\Text[x=9,y=6.65]{$a$}
\end{tikzpicture}
,\label{eq:support2lightcone}
\end{equation}
\begin{equation}
\braket{a_j(t) b_j}=
\begin{tikzpicture}[baseline=(current bounding box.center), scale=0.55]
\foreach \i in {1,2}
{
\Wgategreen{4}{2*\i}
\Wgategreen{3}{1+2*\i}
\Wgategreen{5}{1+2*\i}
\MYcircle{5.5}{1.5+2*\i}
\MYcircle{5.5}{0.5+2*\i}
\MYcircle{2.5}{1.5+2*\i}
\MYcircle{2.5}{0.5+2*\i}
}
\Wgategreen{3}{1}
\Wgategreen{5}{1}
\MYcircle{5.5}{1.5}
\MYcircle{5.5}{0.5}
\MYcircle{2.5}{1.5}
\MYcircle{2.5}{0.5}
\MYsquareB{4}{0}
\draw[very thick] (4.25,0.25) -- (4.5,0.5);
\draw[very thick] (3.75,0.25) -- (3.5,0.5);
\MYsquareB{4}{6}
\draw[very thick] (3.75,5.75) -- (3.5,5.5);
\draw[very thick] (4.25,5.75) -- (4.5,5.5);
\Text[x=4,y=-0.75]{$b$}
\Text[x=4,y=6.75]{$a$}
\end{tikzpicture}
. \label{eq:support2timeaxis}
\end{equation}
%
The correlation function in Eq. (\ref{eq:support2lightcone}) along the light cone is exactly the same as Eq. (\ref{eq:expression_CF_1site}).
By implementing the quantum channel defined as
\begin{equation}
\mathbb{Q}=
\begin{tikzpicture}[baseline=(current  bounding  box.center), scale=0.55]
\Wgategreen{0}{0}
\Wgategreen{-1}{1}
\Wgategreen{1}{1}
\MYcircle{-1.5}{1.5}
\MYcircle{-1.5}{0.5}
\MYcircle{1.5}{1.5}
\MYcircle{1.5}{0.5}
\end{tikzpicture},
\end{equation}
we can express Eq.~\eqref{eq:support2timeaxis} in a compact analytical form $C_{i,i}(t)=\begin{cases}
    \mathrm{Tr}(a\mathbb{Q}^tb), &t=\mathbb{Z},\\
    \mathrm{Tr}(aw\mathbb{Q}^tb), &t=\mathbb{Z}+\frac{1}{2}.\\
\end{cases}$


\begin{figure}[h]
\includegraphics[width=0.9\columnwidth]{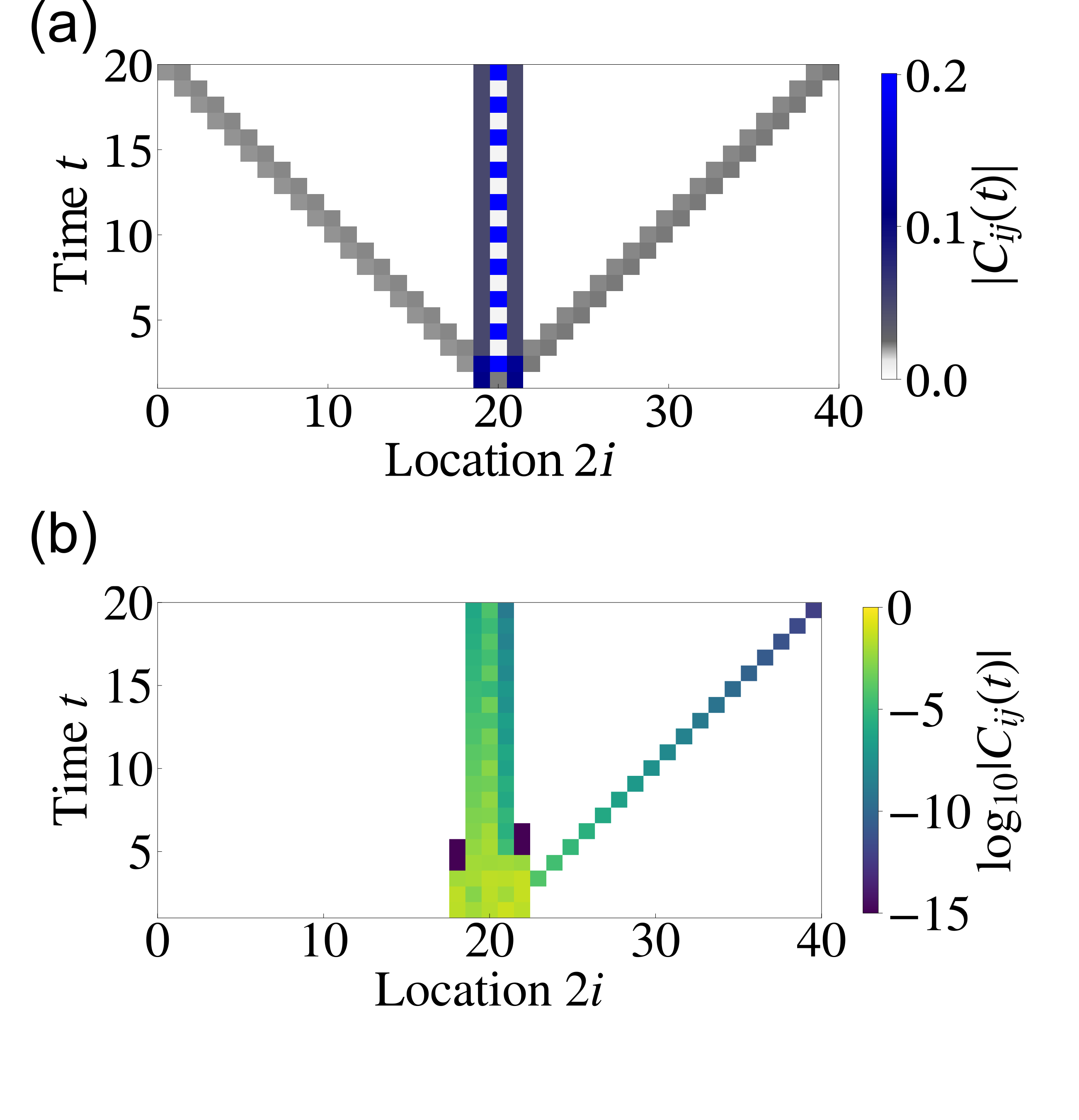}
\caption{(a) The correlation function for $D=6$ dimension, supporting on 2 sites. The gate is a non-ergodic member of $\overline{\mathfrak{L}}_2$ with parametrization given by Eq. (\ref{eq:simplest_2nd_Hierarchy_case}). 
(b) The correlation function for the qubit case supported on 3 sites. The gate is an ergodic element of $\overline{\mathfrak{L}}_2$, with parameters given in the App. \ref{app:parameters}. The asymmetry between the left and right sides results from even/odd effects.
In both figures, $a_{i}=b_{j}=h$, with $h$ a random normalized traceless Hermitian operator. 
$j$ is fixed at $20$. The location of a multi-site operator is defined as the location of its left end. 
}
\label{Correlation_func_sup2}
\end{figure}

Let us have a look at the correlation function's intriguing temporal decay. Generally speaking, the long-term behavior of the correlation function will be dominated by the largest eigenvalue $\lambda$ of the quantum channel, evolving as $\sim \lambda^t$~\footnote{The identity operator is trivially an eigenvector of the quantum channel with eigenvalue $1$, but it does not matter due to the tracelessness of initial operators.}. 
If $|\lambda|=1$, the correlation function will persist without decay. This behavior is referred to as non-ergodic. 
A simple example of $\overline{\mathfrak{L}}_2$ circuit in higher dimension, Eq. (\ref{eq:simplest_2nd_Hierarchy_case}) with $D=6,v_1=v_2=I_D$, falls into this class, as illustrated in Fig. \ref{Correlation_func_sup2}(a). Conversely, if $|\lambda|<1$, the correlation function will be ergodic and exhibit an exponential decay, a point we will come back later in Sec. \ref{Quantum_quench} in the context of quantum quenches. An ergodic example can also be constructed from Eq. (\ref{eq:simplest_2nd_Hierarchy_case}) by choosing $D=6,v_i=\sigma_x\otimes\kappa_i,i\in\{1,2\}$ for almost any $\kappa_i\in\mathbb{SU}(3)$.

\begin{figure}
\includegraphics[width=0.9\columnwidth]{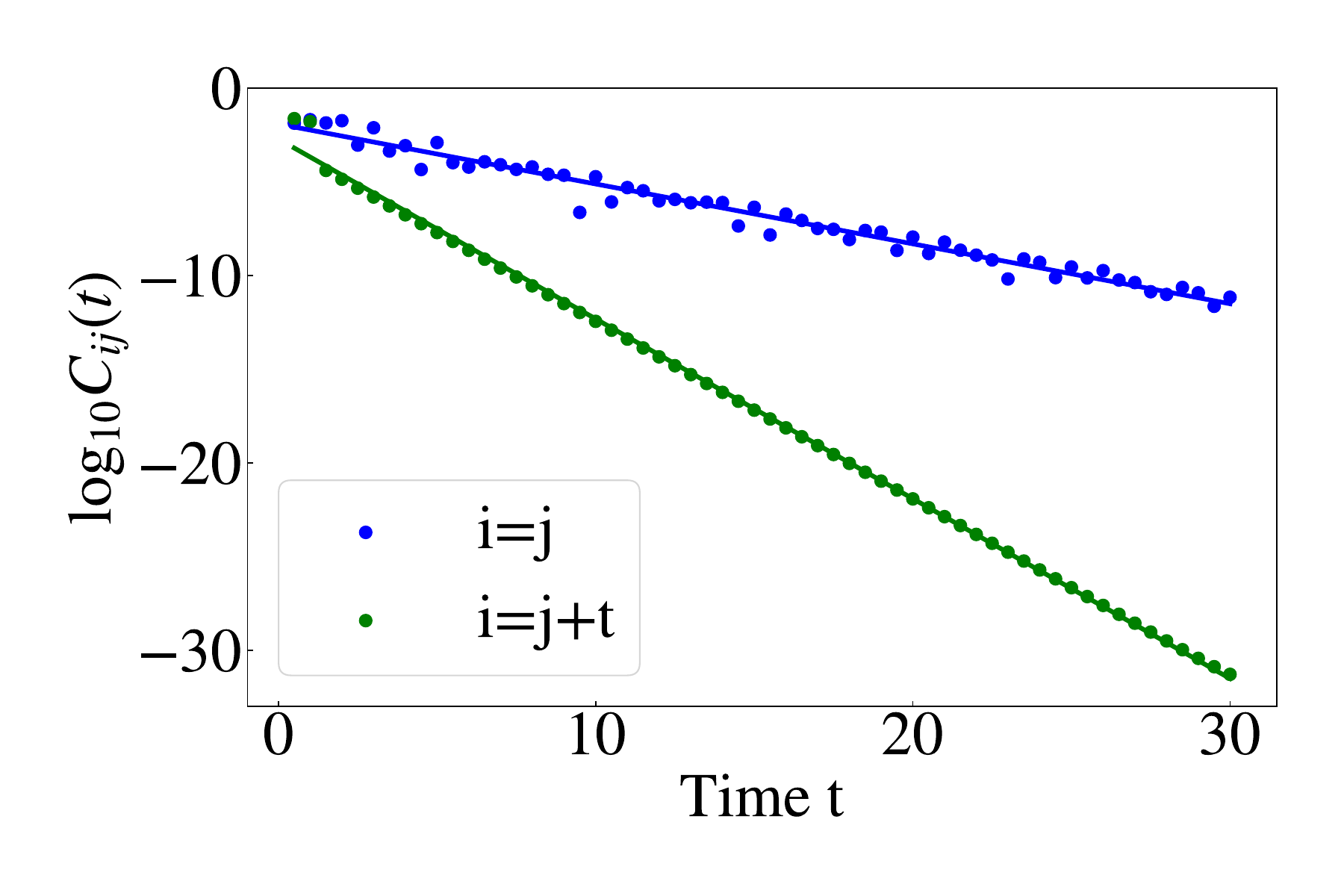}

\caption{This figure demonstrates the exponential decay of the correlation function along both the time axis (i.e., $i=j$) and the light cone (i.e., $i=t+j$). The two-qubit gate from $\overline{\mathfrak{L}}_2$, and the operators $a$ and $b$ are identical to the one used in Fig. \ref{Correlation_func_sup2}(b). The solid line represents a linear fit of the exponential decay.
}
\label{exponential_decay_CF}
\end{figure}

The scenario with three sites operators  follows
 the preceding discussions without any additional difficulty. The correlators along the time axis can be expressed with the quantum channel 
 \begin{equation}
 \mathbb{R}=\begin{tikzpicture}[baseline=(current bounding box.center), scale=0.55]
\Wgategreen{0}{0}
\Wgategreen{2}{0}
\Wgategreen{-1}{1}
\Wgategreen{1}{1}
\MYcircle{2.5}{-0.5}
\MYcircle{2.5}{0.5}
\MYcircle{-1.5}{0.5}
\MYcircle{-1.5}{1.5}
\end{tikzpicture}.
\end{equation}
%
Its largest non-trivial eigenvalue is typically smaller than $1$.
Fig. \ref{Correlation_func_sup2}(b) showcases this ergodic qubit circuit. We further examine the correlators at $i=j$ and $i=j+t$ in Fig. \ref{exponential_decay_CF}. 

Moving beyond the scope of $2-$point correlation functions, $3-$point correlation functions provide more information of the non-equilibrium dynamics. They are defined as follows:
\begin{align}
&C_{i,j,k}(t_{1},t_{2})=\langle a_{i}(t_{1})b_{j}(t_{2})c_{k}\rangle \\
&=D\mathrm{Tr}\left((\mathbb{U}^\dagger)^{t_2}[(\mathbb{U}^\dagger)^{t_1-t_2}a_i\mathbb{U}^{t_1-t_2}b_j]\mathbb{U}^{t_2}c_k\frac{1}{D^{2L}}\right). \notag
\end{align}
If $i$ and $j$ are on the same side of $k$, the $3-$point correlation functions become trivial for $\mathfrak{L}_2$ circuits, i.e., either vanishes or reduces to $2-$point correlations.
Therefore, without loss of generality, we can assume $i<k<j$ such that
\begin{equation}
\begin{aligned}
&C_{i,j,k}(t_1,t_2)=\\
&\begin{tikzpicture}[baseline=(current bounding box.center), scale=0.55]
\Wgategreen{1}{5}
\foreach \i in {0,...,3}
{
\Wgategreen{0}{2*\i}
\Wgategreen{-1}{1+2*\i}
\Wgategreen{1+\i}{3+\i}
}
\foreach \i in {0,1,2}
{
\Wgategreen{-2-\i}{8+\i}
}
\MYcircle{0.5}{6.5}
\foreach \i in {0,...,6}
{
\MYcircle{-1.5}{.5+\i}
}
\foreach \i in {0,1}
{\MYcircle{3.5-\i}{6.5-\i}}
\foreach \i in {0,1,2}
{
\MYcircle{0.5}{\i-0.5}
\MYcircle{-2.5-\i}{7.5+\i}
}
\foreach \i in {0,...,3}
{
\MYcircle{1.5+\i}{2.5+\i}
\MYcircle{-0.5-\i}{7.5+\i}
}
\MYcircle{1.5}{5.5}
\MYcircleB{-0.5}{-0.5}
\MYcircleB{4.5}{6.5}
\MYcircleB{-4.5}{10.5}
\Text[x=-0.5,y=-1] {$c$}
\Text[x=4.5,y=7] {$b$}
\Text[x=-4.5,y=11] {$a$}
\draw[thick, dashed](5.5,6.5)--(7,6.5);
\draw[thick, dashed](5.5,-0.5)--(7,-0.5);
\Text[x=6.25,y=3]{$t_2$};
\draw[thick, dashed](-5.5,10.5)--(-7,10.5);
\draw[thick, dashed](-5.5,-0.5)--(-7,-0.5);
\Text[x=-6.25,y=5]{$t_1$};
\draw[->](6.25,3.5)--(6.25,6);
\draw[->](6.25,2.5)--(6.25,0);
\draw[->](-6.25,5.5)--(-6.25,10);
\draw[->](-6.25,4.5)--(-6.25,0);
\draw[thick, dashed](1,0)--(2,0);
\draw[thick, dashed](1,1)--(2,1);
\Text[x=3,y=0.5]{$t_2-(j-k)-\frac{1}{2}$};
\draw[thick, dashed](-5,0)--(-4,0);
\draw[thick, dashed](-5,6)--(-4,6);
\node at (-4.5,3) [rotate=-90] {$t_1-(k-i)-\frac{1}{2}$};
\draw[thick, dashed](-3,1)--(-2,1);
\draw[thick, dashed](-3,6)--(-2,6);
\node at (-2.5,3.5) {$l$};
\draw[->](-2.5,4)--(-2.5,5.5);
\draw[->](-2.5,3)--(-2.5,1.5);
\end{tikzpicture}
.\label{eq:3pointresult}
\end{aligned}
\end{equation}
Unlike the $\mathfrak{L}_1$, there are non-trivial correlation even when both $a$ and $b$ are strictly inside the light cone.

Nevertheless, despite the fact that the $\mathfrak{L}_2$ condition greatly simplifies the circuit complexity, it does not fully resolve all computational challenges. Viewing from Eq. (\ref{eq:3pointresult}) we obtain that the maximum number of qudits (legs) we need to store when contracting the graph diagonally is $l+\frac{1}{2}$ with $l$ labeled in Eq. (\ref{eq:3pointresult}), thus the computational complexity scales as $e^{\mathcal{O}\left(|(t_{1}-k+i)-(t_{2}-j+k)|\right)}$. 

\subsection{Quantum Quench in $\overline{\mathfrak{L}}_2$ circuits}
\label{Quantum_quench}


In this subsection, we examine the correlation functions for $\overline{\mathfrak{L}}_2$ circuits following a quantum quench, i.e. originating from an initial density matrix $\rho_L(0)$. Here $\rho_L(0)$ can either be a pure state or a mixed state with a local purification.
Therefore we can write it with a local purifying Matrix Product State (MPS) as $\rho_{L}^A(0)=\mathrm{Tr}_{\gamma_{1},\cdots,\gamma_{L}}\ket{\Psi_{L}(A)}\bra{\Psi_{L}(A)}$ 
\cite{kos2023circuits,piroli2020exact}, where
\begin{align}&\ket{\Psi_{L}(A)}  =\\
&\!\!\!\!\!\!\!\!\!\!\!\sum_{\; \; \; \;\; \;\; \; \{i_{k}^{\mathrm{L}},i_{k}^{\mathrm{R}},\gamma_{k}\}} \!\!\!\! \!\! \!\!\!\! \mathrm{Tr}\!\left(A^{(i_{1}^{\mathrm{L}}i_{1}^{\mathrm{R}}\gamma_{1})}\cdots A^{(i_{L}^{\mathrm{L}}i_{L}^{\mathrm{R}}\gamma_{L})}\right)\!\!\ket{i_{1}^{\mathrm{L}}i_{1}^{\mathrm{R}}\gamma_{1}\!\cdots i_{L}^{\mathrm{L}}i_{L}^{\mathrm{R}}\gamma_{L}}.\notag
\end{align}
Here $\gamma_i$ is the purification  index, which we sum over in $\rho_L(0)$.
Without additional specifications, the gates in this subsection are assumed to be from $\overline{\mathfrak{L}}_2$. 
 Graphically, the vectorized density matrix can be represented as 
\begin{equation}
\ket{\rho_L^A(0)}\!=\!\frac{1}{d^L} 
\begin{tikzpicture}[baseline=(current bounding box.center), scale=0.55]
\draw[very thick] (-1.0 ,0.) -- (3.0,0.);
\rhoO{0}{0}\rhoO{2}{0}
\end{tikzpicture}
\!=\!
\begin{tikzpicture}[baseline=(current  bounding  box.center), scale=.6]
\def\dx{0.15}
\def\dy{0.15}
\draw[ thick] (-4.75+.2+\dx,0.+\dy) --(-0.0+0.2+\dx,0.+\dy);
\draw[ thick] (-3.75+\dx,0.75+\dy) -- (-3.75+\dx,-0.0+\dy);
\draw[ thick] (-3.25+\dx,0.75+\dy) -- (-3.25+\dx,0.0+\dy);
\draw[ thick, fill=myblue, rounded corners=2pt] (-4+\dx,0.25+\dy) rectangle (-2.5+\dx,-0.25+\dy);
\draw[ thick] (-1.75+\dx,0.75+\dy) -- (-1.75+\dx,-0.0+\dy);
\draw[ thick] (-1.25+\dx,0.75+\dy) -- (-1.25+\dx,0.0+\dy);
\draw[ thick, fill=myblue, rounded corners=2pt] (-2+\dx,0.25+\dy) rectangle (-0.5+\dx,-0.25+\dy);
\draw[ thick] (-2.75 ,0.25) -- (-2.75,0.25+.2) --(-2.75+\dx,0.25+\dy+.2) --(-2.75+\dx,0.25+\dy);
\draw[ thick] (-4.75 ,0.) -- (-.0+.2,0.);
\draw[ thick] (-3.75,0.75) -- (-3.75,-0.0);
\draw[ thick] (-3.25,-0.0) -- (-3.25,0.75);
\draw[ thick, fill=myorange0, rounded corners=2pt] (-4,0.25) rectangle (-2.5,-0.25);
\draw[ thick] (-1.75,0.75) -- (-1.75,-0.0);
\draw[ thick] (-1.25,-0.0) -- (-1.25,0.75);
\draw[ thick, fill=myorange0, rounded corners=2pt] (-2,0.25) rectangle (-0.5,-0.25);
\draw[ thick] (-0.75 ,0.25) -- (-0.75,0.25+.2) --(-0.75+\dx,0.25+\dy+.2) --(-0.75+\dx,0.25+\dy);
\end{tikzpicture}.
\end{equation}
A physical density matrix is normalized in the thermodynamic limit
\begin{equation}
\lim_{L\to\infty}\braket{I|\rho_L^A(0)}=\lim_{L\to\infty}\mathrm{Tr}E(0)^L=1,
\end{equation}
where $E(0)$ is the space transfer matrix
\begin{equation}
E(0)=\begin{tikzpicture}[baseline=(current bounding box.center), scale=0.55]
\rhoO{0}{0}
\MYcircle{-0.5}{0.5}
\MYcircle{0.5}{0.5}
\end{tikzpicture}.
\end{equation}
This implies that $E(0)$ has a unique non-degenerate left and right fixed point whose eigenvalue is one 
\begin{equation}
    \lim_{L\to\infty}E(0)^L=\ket{\square}\bra{\vartriangle}=
     \begin{tikzpicture}[baseline=(current bounding box.center), scale=0.55]
\draw[very thick] (0,0) -- (0.5,0);
\MYsquare{0.5}{0};
\draw[very thick] (1.0,0) -- (1.5,0);
\MYtriangle{1.0}{0}
\end{tikzpicture}
\label{eq:leftrightfixedpoint}
\end{equation}
with $\braket{\vartriangle|\square}=1$.
\subsubsection{$1-$point correlation functions}
The most basic information about the dynamic from a quantum quench is contained in the $1-$point correlation function, which specifies the relaxation and thermalization of local observables. It is defined as
\begin{equation}
\lim_{L\to\infty}\braket{O_1|\rho_L(t)}=\lim_{k\to\infty}\mathrm{Tr}\left(E^{k}(t)E_{O_{1}}(t)E^{k}(t)\right).
\end{equation}
Here, $E(t)=E_{\1}(t)$ and $E_{O_{1}}(t)$ are appropriate space transfer matrices
\begin{equation}
\begin{aligned}
\braket{O_1|\rho_L(t)} & =
\begin{tikzpicture}[baseline=(current bounding box.center), scale=0.55]
\draw [very thick] (-0.5,0) -- (6.5,0);
\foreach \i in {0,2,4,6}
{\rhoO{\i}{0}}
\foreach \i in {1,3,5}
{
\foreach \j in {1,3}
{
\Wgategreen{\i}{\j}
}
}
\foreach \i in {0,2,4,6}
{
\foreach \j in {2,4}
{
\Wgategreen{\i}{\j}
}
}
\foreach \i in {0,1,2,3,5,6,7}
{
\MYcircle{\i-0.5}{4.5}
}
\MYcircleB{3.5}{4.5}
\Text[x=3.5,y=5] {$O_1$};
\draw[gray, dashed](2.7,4.65)--(2.7,-0.15)--(4.7,-0.15)--(4.7,4.65)--cycle;
\Text[x=3.5,y=-.65] {$E_{O_1}$(t)};
\end{tikzpicture}
\end{aligned}
.
\end{equation}
In this subsection we always assume periodic boundary conditions with large enough $L$ such that the transfer matrix can be replaced by its fixed point. 

If the initial state satisfies the following \emph{$1-$point solvable
condition for $\overline{\mathfrak{L}}_2$}
\begin{equation}
\begin{tikzpicture}[baseline=(current bounding box.center), scale=0.55]
\rhoO{0}{0}
\Wgategreen{-1}{1}
\MYcircle{-0.5}{1.5}
\MYcircle{0.5}{0.5}
\MYsquare{0.5}{0}
\end{tikzpicture}
=
\begin{tikzpicture}[baseline=(current bounding box.center), scale=0.55]
\draw[very thick] (-0.5,0) -- (0.5,0);
\Wgategreen{-1}{1}
\MYcircle{-0.5}{1.5}
\MYcircle{-0.5}{0.5}
\MYsquare{0.5}{0}
\end{tikzpicture}
, \label{eq:transfer_condi}
\end{equation}
with \begin{tikzpicture}[baseline=(current bounding box.center), scale=0.55]
\draw[very thick] (0,0) -- (0.5,0);
\MYsquare{0.5}{0}
\end{tikzpicture} being the right eigenvector of $E(0)$, \footnote{Note that by contracting the left top leg with an empty bullet in both sides of Eq. (\ref{eq:transfer_condi}), we can directly show that Eq. (\ref{eq:transfer_condi}) implies \begin{tikzpicture}[baseline=(current bounding box.center), scale=0.55]
\draw[very thick] (0,0) -- (0.5,0);
\MYsquare{0.5}{0}
\end{tikzpicture} is the right eigenvector of $E(0)$.}
we can find an eigenstate of the transfer matrix with eigenvalue $1$ 
\begin{equation}
\begin{tikzpicture}[baseline=(current bounding box.center), scale=0.55]
\rhoO{-1}{0}
\foreach \j in {1,3}
{
\Wgategreen{-2}{\j}
}
\foreach \j in {2,4}
{
\Wgategreen{-1}{\j}
}
\MYcircle{-1.5}{4.5}
\MYcircle{-0.5}{4.5}
\draw[very thick] (-0.5,0)--(0.5,0);
\MYsquare{0.5}{0}
\foreach \j in {1,3}
{
\Wgategreen{0}{\j}
}
\foreach \j in {1,2,3,4}
{
\MYcircle{0.5}{\j-0.5}
}
\draw[gray, dashed] (-2.5,-0.15)--(-0.35,-0.15)--(-0.35,4.65)--(-2.5,4.65)--cycle;
\Text[x=-1.5,y=-.65]{$E(t)$};
\end{tikzpicture}
\;
=
\;
\begin{tikzpicture}[baseline=(current bounding box.center), scale=0.55]
\draw[very thick] (-0.5,0)--(0.5,0);
\MYsquare{0.5}{0}
\foreach \j in {1,3}
{
\Wgategreen{0}{\j}
}
\foreach \j in {1,2,3,4}
{
\MYcircle{0.5}{\j-0.5}
}
\end{tikzpicture}
.
\end{equation}
Since $\lim_{L\to\infty}\mathrm{Tr}\left(\rho_L(t)\right)=1$ due
to the normalization, the largest eigenvalue of the transfer matrix
must be $1$ and non-degenerate. Therefore, the $1-$point correlation
function can be analytically calculated with this eigenvector.

It is worth to note that if the initial state satisfies the condition 
$\begin{tikzpicture}[baseline=(current bounding box.center), scale=0.55]
\rhoO{0}{0}
\MYcircle{0.5}{0.5}
\MYsquare{0.5}{0}
\end{tikzpicture}
\;
=
\begin{tikzpicture}[baseline=(current bounding box.center), scale=0.55]
\draw[very thick] (-0.5,0) -- (0.5,0);
\draw[very thick] (-0.5,0.5) -- (0.5,0.5);
\MYcircle{0.5}{0.5}
\MYsquare{0.5}{0}
\end{tikzpicture}
\,
,$ Eq. (\ref{eq:transfer_condi}) is automatically satisfied. This implies
that we have identified a larger solvable class than both the pure solvable initial states for dual-unitary evolution~\cite{piroli2020exact} and in general mixed initial states for open 3-way unital evolution~\cite{kos2023circuits}.

The correlation function for $2-$site observables after a quench is very similar to Eq. (\ref{eq:support2timeaxis}). The only difference is the substitution of $\begin{tikzpicture}[baseline=(current  bounding  box.center), scale=0.45]
\MYsquareB{0}{0}
\draw[very thick](0.25,0.25)--(0.5,0.5);
\draw[very thick](-0.25,0.25)--(-0.5,0.5);
\end{tikzpicture}$ at the base for 
$\begin{tikzpicture}[baseline=(current  bounding  box.center), scale=0.55]
\rhoO{0}{0}
\MYsquare{0.5}{0}
\MYtriangle{-0.5}{0}
\end{tikzpicture}$.
\begin{equation}
\braket{O_1|\rho_L(t)} =\begin{tikzpicture}[baseline=(current bounding box.center), scale=0.55]
\foreach \i in {1,2}
{
\Wgategreen{4}{2*\i}
\Wgategreen{3}{1+2*\i}
\Wgategreen{5}{1+2*\i}
\MYcircle{5.5}{1.5+2*\i}
\MYcircle{5.5}{0.5+2*\i}
\MYcircle{2.5}{1.5+2*\i}
\MYcircle{2.5}{0.5+2*\i}
}
\Wgategreen{3}{1}
\Wgategreen{5}{1}
\MYcircle{5.5}{1.5}
\MYcircle{5.5}{0.5}
\MYcircle{2.5}{1.5}
\MYcircle{2.5}{0.5}
\rhoO{4}{0}
\MYsquare{4.5}{0}
\MYtriangle{3.5}{0}
\draw[very thick] (4.25,0.25) -- (4.5,0.5);
\draw[very thick] (3.75,0.25) -- (3.5,0.5);
\MYsquareB{4}{6}
\draw[very thick] (3.75,5.75) -- (3.5,5.5);
\draw[very thick] (4.25,5.75) -- (4.5,5.5);
\Text[x=4,y=6.75]{$O_1$}
\end{tikzpicture},
\end{equation}
where $\bra{\vartriangle}$ and $\ket{\square}$ are the left and right fixed points from Eq. (\ref{eq:leftrightfixedpoint}). Following the same argument as in Sec. \ref{subsec:Biggersupports}, the long time behavior is dictated by the largest eigenvalue of the quantum channel, $\mathbb{Q}$. The eigenspectrum of $\mathbb{Q}$ can be completely deduced for qubits. Using the general parametrization of gates from $\overline{\mathfrak{L}}_2$ in accordance with Eqs. (\ref{eq:parameter2qubit}) and (\ref{eq:condition2ndqubit}), whereby $v_{3}=v_{4}=I$, $\theta_{i}=\arcsin{\frac{1}{\sqrt{2}\sin{r_{i}}}},i\in\{1,2\}$, only two non-zero eigenvalues, $\{1,\lambda\}$, exist. Eigenvalue $1$ corresponds to the trivial eigenvector, i.e. identity and $\lambda$ is given by
\begin{equation}
\begin{aligned}
&\lambda=\cos^2{(\phi_2-\phi_1)}\\
&-2(\sqrt{-\cos{2r_2}}\cos{r_1}+\sqrt{-\cos{2r_1}}\cos{r_2})^2.
\label{eq:quantumquenchchannel2qubit}
\end{aligned}
\end{equation}
Here $r_1,r_2$ are real numbers from the interval $[\frac{\pi}{4},\frac{3\pi}{4}]$.

The dynamics is non-ergodic with $\lambda=1$ if $\phi_1=\phi_2+0,\pi$ and either $r_1=\pi-r_2$ or $\cos{2r_1}=\cos{2r_2}=0$. On the other hand, the circuit is non-ergodic also with $\lambda=-1$ if $\cos(\phi_1-\phi_2)=0$ and the last term in Eq. (\ref{eq:quantumquenchchannel2qubit}) equals $1$, for example, by $r_1=\frac{\pi}{2},r_2=\frac{\pi}{4}$. This gives all possible non-ergodic circuits for two-qubit gates from $\overline{\mathfrak{L}}_2$, apart from a tensor product of single-site gates. All other examples are ergodic and show exponential decay.

\begin{figure}
\includegraphics[width=0.9\columnwidth]{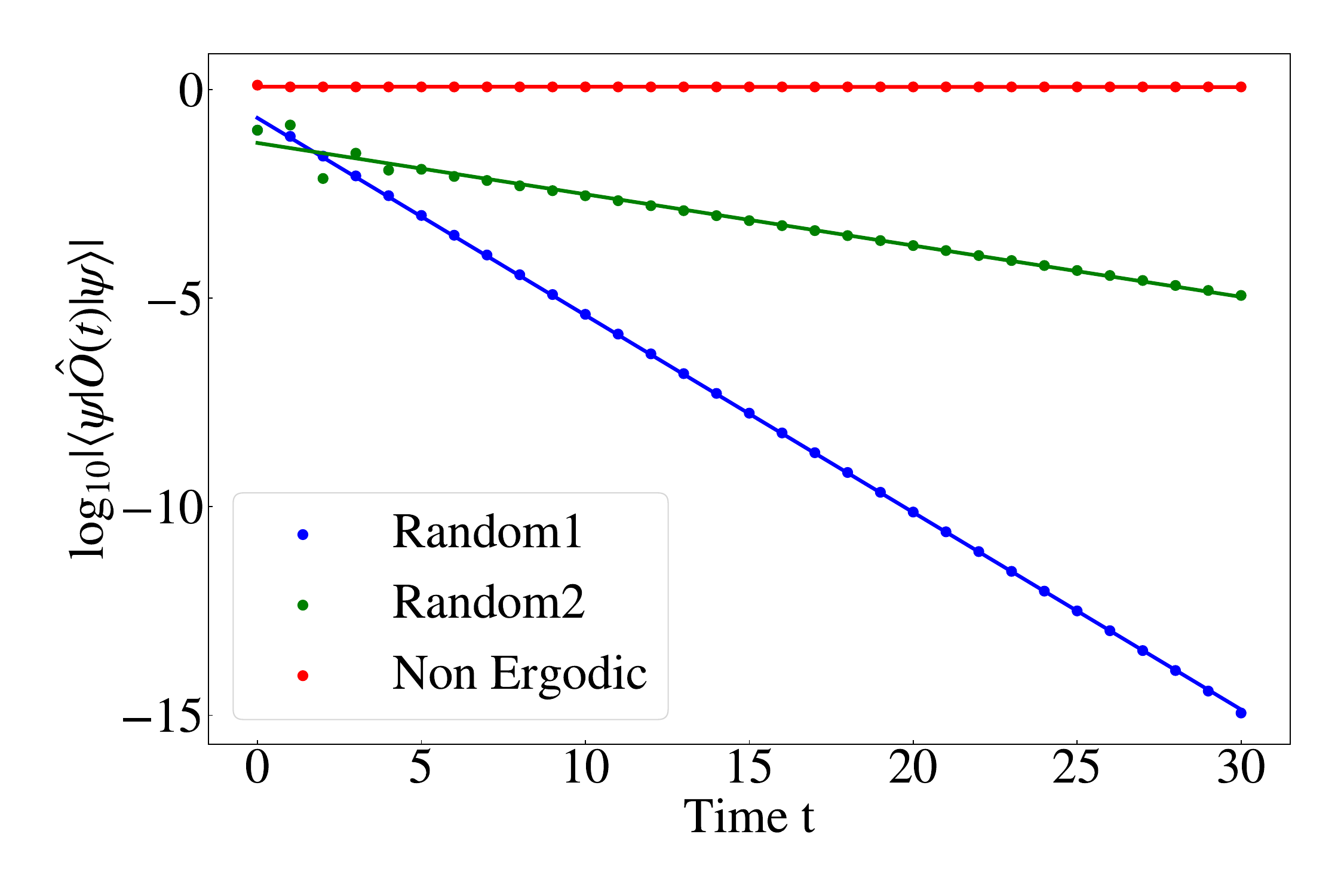}

\caption{The correlation function of a random two-site observable following a quantum quench starting from the Bell state $\ket{\Psi_L}$ defined in Eq. (\ref{eq:manybodyinitialstate}). The dynamics is governed by a $\overline{\mathfrak{L}}_2$ circuit (symbols). We show its linear fit by solid lines.
The blue and green data points show an exponential decay of the expectation value of observable, indicative of thermalization dynamics. In contrast, the red points remains constant at long times implying non-thermalization and non-ergodicity. The parameters for these three circuits can be found in App.~\ref{app:parameters}.
}
\label{quench_correlation}
\end{figure}

The most straightforward solution of initial states from Eq. (\ref{eq:transfer_condi}) is the pure state with a bond dimension $1$. One example is the qubit Bell state
\begin{equation}
\ket{\Psi_L}=\otimes_{k=1}^L\frac{\ket{01}_{2k-1,2k}+\ket{10}_{2k-1,2k}}{\sqrt{2}},
\label{eq:manybodyinitialstate}
\end{equation}
which we use in Fig.~\ref{quench_correlation}, where we show both thermalizing and non-thermalizing behaviors.
In contrast to the situation here, for $\mathfrak{L}_1$ (dual-unitarity) the expectation values of local observables trivially vanish at all times, even for clearly non-ergodic circuits such as circuits made out of SWAPs.
%

\subsubsection{$2-$point correlation functions}
Finally, we are moving to the $2-$point correlation functions after a quench, which are non trivial even for the $\mathfrak{L}_1$~\cite{piroli2020exact}.
Diagrammatically they  are represented as
\begin{align}\label{eq:quench2}
&C_{ij}(t) = 
\begin{tikzpicture}[baseline=(current  bounding  box.center), scale=0.55]
\foreach \i in {0,...,1}
{
\Wgategreen{-2}{0+2*\i}
\Wgategreen{0}{0+2*\i}
\Wgategreen{4}{0+2*\i}
\Wgategreen{6}{0+2*\i}
\Wgategreen{-1}{1+2*\i}
\Wgategreen{1}{1+2*\i}
\Wgategreen{5}{1+2*\i}
\Wgategreen{7}{1+2*\i}
}
\Text[x=2.5,y=1.5]{$\cdots$};
\Text[x=2.5,y=-1]{$\cdots$};
\draw[very thick] (-2.25,-1) -- (2,-1);
\draw[very thick] (3,-1) -- (7.5,-1);
\foreach \i in {2.5,4.5,8.5,10.5}
{\rhoO{-3.5+\i}{-1}}
\foreach \i in {-1,0,1,2,5,6,7,8}
{ 
\draw[thick, fill=white] (\i-0.5,4-0.5) circle (0.1cm);
}
\MYcircleB{7.5}{3.5}
\MYcircleB{-1.5}{3.5}
\Text[x=-1.5,y=4.0]{$a$}
\Text[x=7.5,y=4.0]{$b$}
\end{tikzpicture}
.
\end{align} 

By leveraging time unitarity, we can deduce the emergence of two backward propagating light cones originating from the operators, interconnected by the initial state's transfer matrix $E(0)$
\begin{widetext}
\begin{align}
C_{ij}(t)=
\begin{tikzpicture}[baseline=(current  bounding  box.center), scale=0.55]
\Wgategreen{-8}{0}
\Wgategreen{-6}{0}
\Wgategreen{-4}{0}
\Wgategreen{-2}{0}
\Wgategreen{4}{0}
\Wgategreen{6}{0}
\Wgategreen{8}{0}
\Wgategreen{10}{0}
\Wgategreen{-5}{3}
\Wgategreen{7}{3}
\Wgategreen{-6}{2}
\Wgategreen{-4}{2}
\Wgategreen{6}{2}
\Wgategreen{8}{2}
\Wgategreen{-7}{1}
\Wgategreen{-5}{1}
\Wgategreen{-3}{1}
\Wgategreen{5}{1}
\Wgategreen{7}{1}
\Wgategreen{9}{1}
\draw[very thick] (-10.5,-1) -- (13.5,-1);
\draw[very thick,dotted] (-12.5,-1) -- (-10.5,-1);
\draw[very thick,dotted] (13.5,-1) -- (14.5,-1);
\foreach \i in {-7.5,-5.5,-3.5,-1.5,0.5,2.5,4.5,6.5,8.5,10.5,12.5,14.5,16.5}
{ \rhoO{-3.5+\i}{-1} }
\foreach \i in {0,...,3}
{
\MYcircle{\i-4.5}{3.5-\i}
\MYcircle{\i+4-.5}{0.5+\i}
\MYcircle{\i-.5+8}{3.5-\i}
\MYcircle{\i-4-4.5}{0.5+\i}
}
\MYcircle{-9.5}{-.5}  
\MYcircle{-10.5}{-.5} 
\MYcircle{-11.5}{-.5}
\MYcircle{11.5}{-.5}  
\MYcircle{12.5}{-.5}  
\MYcircle{13.5}{-.5} 
\MYcircle{2.5}{-0.5}
\MYcircle{1.5}{-0.5}
\MYcircle{0.5}{-0.5}
\MYcircle{-0.5}{-0.5}
\MYcircleB{7.5}{3.5} 
\MYcircleB{-5.5}{3.5}
\Text[x=-5.5,y=4.0]{$a$}
\Text[x=7.5,y=4.0]{$b$}
\end{tikzpicture}.
\label{eq:quench4}
\end{align}
\end{widetext} 
The number of repetitions of the transfer matrices $E(0)$ in the central region
depends on the separation, given by $j-i-(t_{1}+t_{2})+\frac{1}{2}$. The suitable 
\emph{$\mathfrak{L}_2$ 2-point correlator solvability condition} for this correlation function is 
\begin{equation}
\begin{tikzpicture}[baseline=(current  bounding  box.center), scale=0.55]
\rhoO{0}{0}
\MYsquare{0.5}{0}
\MYcircle{0.5}{0.5}
\end{tikzpicture}
=
\begin{tikzpicture}[baseline=(current  bounding  box.center), scale=0.6]
\draw[very thick] (-0.5,0) -- (0.5,0);
\draw[very thick] (-0.5,0.5) -- (0.5,0.5);
\MYsquare{0.5}{0}
\MYcircle{0.5}{0.5}
\end{tikzpicture}\;
;
\qquad
\begin{tikzpicture}[baseline=(current  bounding  box.center), scale=0.6]
\rhoO{0}{0}
\MYsquare{0.5}{0}
\Wgategreen{1}{1}
\MYcircle{1.5}{1.5}
\MYcircle{1.5}{0.5}
\end{tikzpicture}
=
\begin{tikzpicture}[baseline=(current  bounding  box.center), scale=0.55]
\draw[very thick] (-0.5,0) -- (0.5,0);
\draw[very thick] (-0.5,0.5) -- (0.5,0.5);
\draw[very thick] (-0.5,1.5) -- (0.5,1.5);
\MYsquare{0.5}{0}
\MYcircle{0.5}{0.5}
\MYcircle{0.5}{1.5}
\end{tikzpicture}\;
,\label{eq:solvable2points}
\end{equation}
and similar expression from left
\begin{equation}
\begin{tikzpicture}[baseline=(current  bounding  box.center), scale=0.55]
\rhoO{0}{0}
\MYtriangle{-0.5}{0}
\MYcircle{-0.5}{0.5}
\end{tikzpicture}
=
\begin{tikzpicture}[baseline=(current  bounding  box.center), scale=0.6]
\draw[very thick] (-0.5,0) -- (0.5,0);
\draw[very thick] (-0.5,0.5) -- (0.5,0.5);
\MYtriangle{-0.5}{0}
\MYcircle{-0.5}{0.5}
\end{tikzpicture}\;
;
\qquad
\begin{tikzpicture}[baseline=(current  bounding  box.center), scale=0.6]
\rhoO{0}{0}
\MYtriangle{-0.5}{0}
\Wgategreen{-1}{1}
\MYcircle{-1.5}{1.5}
\MYcircle{-1.5}{0.5}
\end{tikzpicture}
=
\begin{tikzpicture}[baseline=(current  bounding  box.center), scale=0.55]
\draw[very thick] (-0.5,0) -- (0.5,0);
\draw[very thick] (-0.5,0.5) -- (0.5,0.5);
\draw[very thick] (-0.5,1.5) -- (0.5,1.5);
\MYtriangle{-0.5}{0}
\MYcircle{-0.5}{0.5}
\MYcircle{-0.5}{1.5}
\end{tikzpicture}\;
.\label{eq:solvable2pointsleft}
\end{equation}
Here $\ket{\vartriangle}$ and $\ket{\square}$ are the left and right unique fixed point from Eq. (\ref{eq:leftrightfixedpoint}). 
Note that this condition is stronger than the solvability condition for the 1-point correlators. 
Following the same argument in~\cite{kos2023circuits}, it can be shown that any MPDO in the local purification form fulfilling solvability condition can be cast in right canonical form, which we use.
Thus \begin{tikzpicture}[baseline=(current  bounding  box.center), scale=0.55]
\draw[very thick] (-0.,0) -- (0.5,0);
\MYsquare{0.5}{0}
\end{tikzpicture} is replaced by \begin{tikzpicture}[baseline=(current  bounding  box.center), scale=0.55]
\draw[very thick] (-0.,0) -- (0.5,0);
\MYcircle{0.5}{0}
\end{tikzpicture} (vectorized identity).

Supposing that the length $L$ is large enough, we can replace the outer parts of Eq. (\ref{eq:quench4}) by Eq. (\ref{eq:leftrightfixedpoint}).
After this simplification, we obtain
\begin{widetext}
\begin{align}
C_{ij}(t)=
\begin{tikzpicture}[baseline=(current  bounding  box.center), scale=0.55]
\Wgategreen{-8}{0}
\Wgategreen{-6}{0}
\Wgategreen{-4}{0}
\Wgategreen{-2}{0}
\Wgategreen{4}{0}
\Wgategreen{6}{0}
\Wgategreen{8}{0}
\Wgategreen{10}{0}
\Wgategreen{-5}{3}
\Wgategreen{7}{3}
\Wgategreen{-6}{2}
\Wgategreen{-4}{2}
\Wgategreen{6}{2}
\Wgategreen{8}{2}
\Wgategreen{-7}{1}
\Wgategreen{-5}{1}
\Wgategreen{-3}{1}
\Wgategreen{5}{1}
\Wgategreen{7}{1}
\Wgategreen{9}{1}
\draw[very thick] (-8.5,-1) -- (11.5,-1);
\foreach \i in {-5.5,-3.5,-1.5,0.5,2.5,4.5,6.5,8.5,10.5,12.5,14.5}
{ \rhoO{-3.5+\i}{-1} }
\foreach \i in {0,...,3}
{
\MYcircle{\i-4.5}{3.5-\i}
\MYcircle{\i+4-.5}{0.5+\i}
\MYcircle{\i-.5+8}{3.5-\i}
\MYcircle{\i-4-4.5}{0.5+\i}
}
\MYcircle{-9.5}{-.5}  
\MYcircle{11.5}{-.5}  
\MYcircle{2.5}{-0.5}
\MYcircle{1.5}{-0.5}
\MYcircle{0.5}{-0.5}
\MYcircle{-0.5}{-0.5}
\MYcircleB{7.5}{3.5} 
\MYcircleB{-5.5}{3.5}
\Text[x=-5.5,y=4.0]{$a$}
\Text[x=7.5,y=4.0]{$b$}
\MYtriangle{-9.5}{-1}
\MYcircle{11.5}{-1}
\end{tikzpicture}.
\label{eq:quench5}
\end{align}
\end{widetext} 
The leftmost and rightmost corners can be further reduced by the first equations from (\ref{eq:solvable2points}) and (\ref{eq:solvable2pointsleft}). Then with the help of the second equations from (\ref{eq:solvable2points}), (\ref{eq:solvable2pointsleft}) and the contraction properties of the $\overline{\mathfrak{L}}_2$, we arrive at 
\begin{align}
&C_{ij}(t)= \nonumber\\
&\begin{tikzpicture}[baseline=(current  bounding  box.center), scale=0.55]
\Wgategreen{-4}{0}
\Wgategreen{-2}{0}
\Wgategreen{4}{0}
\Wgategreen{6}{0}
\Wgategreen{-5}{3}
\Wgategreen{7}{3}
\Wgategreen{-4}{2}
\Wgategreen{6}{2}
\Wgategreen{-5}{1}
\Wgategreen{-3}{1}
\Wgategreen{5}{1}
\Wgategreen{7}{1}
\draw[very thick] (-5,-1) -- (7,-1);
\foreach \i in {-1.5,0.5,2.5,4.5,6.5,8.5,10.5}
{ \rhoO{-3.5+\i}{-1} }
\foreach \i in {0,...,3}
{
\MYcircle{\i-4.5}{3.5-\i}
\MYcircle{\i+4-.5}{0.5+\i}
}
\foreach \i in {0,1,2,3}
{
\MYcircle{-5.5}{\i-0.5}
\MYcircle{7.5}{\i-0.5}
}
\MYtriangle{-5.5}{-1}
\MYcircle{7.5}{-1}
\MYcircle{2.5}{-0.5}
\MYcircle{1.5}{-0.5}
\MYcircle{0.5}{-0.5}
\MYcircle{-0.5}{-0.5}
\MYcircleB{7.5}{3.5} 
\MYcircleB{-5.5}{3.5}
\Text[x=-5.5,y=4.0]{$a$}
\Text[x=7.5,y=4.0]{$b$}
\end{tikzpicture}.
\label{eq:quench4_2}
\end{align}
Applying the first equation from (\ref{eq:solvable2points}), we finally
obtain a considerably simplified expression
\begin{align}
&C_{ij}(t)=\nonumber\\
&\begin{tikzpicture}[baseline=(current  bounding  box.center), scale=0.55]
\draw[very thick](-4,0)--(0,0);
\rhoO{0}{0}
\rhoO{-2}{0}
\rhoO{-4}{0}
\foreach \i in {1,2,3,4}
{
\Wgategreen{\i}{\i}
\Wgategreen{-\i-4}{\i}
\MYcircle{\i-0.5}{\i+0.5}
\MYcircle{\i+0.5}{\i-0.5}
\MYcircle{0.5-\i-4}{\i+0.5}
\MYcircle{-0.5-\i-4}{\i-0.5}
}
\foreach \i in {0,1,2,3}
{
\MYcircle{-\i-0.5}{0.5}
}
\MYcircle{0.5}{0}
\MYtriangle{-4.5}{0}
\MYcircleB{4.5}{4.5}
\MYcircleB{-8.5}{4.5}
\Text[x=4.5,y=5]{$b$}
\Text[x=-8.5,y=5]{$a$}
\end{tikzpicture}.
\end{align}
Interestingly, if the separation $j-i$ is shorter than $t_{1}+t_{2}$,
the correlation function will vanish for traceless operators, which is the same as for $\mathfrak{L}_1$~\cite{piroli2020exact}.

Solvability condition from Eq. (\ref{eq:solvable2points}) can also be formulated in an algebraic form.
If we define an operator $K_{\gamma}$ as
\begin{equation}
K^{\dagger}_{\gamma}:=\sum_{i^{\mathrm{L}}j^{\mathrm{R}}}A^{(i^{\mathrm{L}}j^{\mathrm{R}}\gamma)}\otimes|i^{\mathrm{L}}\rangle\langle j^{\mathrm{R}}|,
\end{equation} 
the solvability condition for an initial state in the right canonical form can be written as
\begin{equation}
\begin{aligned}\sum_{\gamma}K_{\gamma}^{\dagger}K_{\gamma} & =\frac{I_{d\chi}}{d};\\
\sum_{\gamma}I_{d}\otimes K_{\gamma}^{\dagger}(\tilde{U^{\dagger}}\tilde{U}\otimes I_{\chi})K_{\gamma}\otimes I_{d} & =\frac{I_{d^{2}\chi}}{d}.
\end{aligned}
\end{equation}
The same discussion holds from the left side, except that we need to keep $\ket{\vartriangle}$ instead of $\ket{\mcirc}$. 

For example, where the initial state is pure, that is, without summation
over $\gamma$, the first equation implies that $K$ is proportional
to a unitary operator. This leads to the conclusion that $U$ is a $\mathfrak{L}_1$, as per the second equation. Consequently, a solvable initial
state for $\overline{\mathfrak{L}}_2$ can only be a mixed state. As an example, if we choose $U=\mathrm{CNOT}$ and the bond dimension for
MPS to be $1$, a nontrivial solution is $K_{0}=\frac{I}{2\sqrt{2}},K_{1}=\frac{\sigma_{Y}}{2\sqrt{2}}$.



\section{Conclusions and perspectives}
\label{sec:conclusions}

In this paper, we have generalized dual-unitary circuits to larger solvable classes referred to as levels of the Hierarchy. These families of circuits with comparatively relaxed conditions exhibit richer physical properties than the dual-unitary ones. Most intriguingly, their correlation functions are non-zero at the initial site and later times. Furthermore, these circuits show  nontrivial thermalization of local operators, in contrast with the standard dual-unitary circuit. 

In our work we provide the complete parametrization of $\mathfrak{L}_1,\mathfrak{L}_2,\mathfrak{L}_3$ for qubit circuits. For local dimensions bigger than two, we propose a systematic approach using the Clifford group to construct novel examples, which include both ergodic and nonergodic instances.

It is worth mentioning that even though we focus on the translational invariant dynamics, the results and methods presented do not depend crucially on this property and can be extended to time and space inhomogeneous settings. 
This study, therefore, provides a strong foundation for further explorations in the realm of solvable models and their real-world applications. 
They can be directly implemented on current NISQ devices, and used as benchmarks in the realms where ordinary classical simulations are no longer possible.

We mentioned that one interesting example is the CNOT gate and its extensions. They are a particular cases of Floquet East model, which shows interesting physics, e.g. hydrophobicity and localization, both in classical~\cite{klobas2023exact} and quantum realm~\cite{bertini2023localised}.
Interestingly, for a subset of the solvable $\mathfrak{L}_2$ connected to CNOT,  exact solutions can also be obtained using the so called zipper equations~\cite{klobas2021exact} by devising an exact eigenvectors of the transfer matrix~\cite{unpublished}. 

Of course, the ideas presented here can be extended to conditions on bunch of gates in different configurations from the ones presented here.
One direct way is generalizing the dual-unitarity
round-a-face~\cite{prosen2021manybody}. It is feasible to define the $\mathfrak{L}_2$ condition in this context as
\begin{equation}
\begin{tikzpicture}[baseline=(current  bounding  box.center), scale=0.55]
\draw[very thick, fill=mygreen,rounded corners=2pt](0,0.5)--(0.5,0)--(0,-0.5)--(-0.5,0)--cycle;
\draw[very thick,fill=mygreen,rounded corners=2pt](-0.5,0)--(0,-0.5)--(-0.5,-1)--(-1,-0.5)--cycle;
\MYcircle{0}{0.5}
\MYcircle{0.5}{0}
\MYcircle{0}{-0.5}
\MYcircle{-0.5}{-1}
\end{tikzpicture}
=
\begin{tikzpicture}[baseline=(current  bounding  box.center), scale=0.55]
\draw[very thick,fill=mygreen,rounded corners=2pt](-0.5,0)--(0,-0.5)--(-0.5,-1)--(-1,-0.5)--cycle;
\draw[very thick](-1,0.5)--(-0.5,0.5);
\MYcircle{-0.5}{0.5}
\MYcircle{-0.5}{0}
\MYcircle{0}{-0.5}
\MYcircle{-0.5}{-1}
\end{tikzpicture}.
\end{equation}
We anticipate these circuits to exhibit similar features as the $\mathfrak{L}_2$ circuits discussed here. For examples, their correlation functions might exclusively exist along three directions. However, parametrization of these types of circuits and their subsequent physical applications represent an intriguing path for future research.

One can also be  more bold, and demand for instance the following conditions:
	\begin{eqnarray}
		\begin{tikzpicture}[baseline=(current  bounding  box.center), scale=0.7]
			\Wgategreen{0}{1}
			\Wgategreen{1}{0}
			\Wgategreen{0}{-1}
			\Wgategreen{-1}{0}
			\Wgategreen{1}{-2}
			\MYcircle{1.5}{0.5}
			\MYcircle{1.5}{-0.5}
			\MYcircle{1.5}{-1.5}
			\MYcircle{1.5}{-2.5}
			\MYcircle{0.5}{1.5}
		\end{tikzpicture}&=&
		\begin{tikzpicture}[baseline=(current  bounding  box.center), scale=0.7]
	\Wgategreen{0}{1}
\Wgategreen{0}{-1}
\Wgategreen{-1}{0}
\Wgategreen{1}{-2}
\MYcircle{.5}{0.5}
\MYcircle{.5}{-0.5}
\MYcircle{1.5}{-1.5}
\MYcircle{1.5}{-2.5}
\MYcircle{0.5}{1.5}
		\end{tikzpicture},\nonumber\\
		\begin{tikzpicture}[baseline=(current  bounding  box.center), scale=0.7]
		\Wgategreen{0}{-1}
		\Wgategreen{-1}{0}
		\Wgategreen{0}{1}
		\Wgategreen{1}{0}
		\Wgategreen{-1}{2}
		\MYcircle{-1.5}{-0.5}
		\MYcircle{-1.5}{0.5}
		\MYcircle{-1.5}{1.5}
		\MYcircle{-1.5}{2.5}
		\MYcircle{-0.5}{-1.5}
		\end{tikzpicture}&=&
		\begin{tikzpicture}[baseline=(current  bounding  box.center), scale=0.7]
		\Wgategreen{0}{-1}
		\Wgategreen{0}{1}
		\Wgategreen{1}{0}
		\Wgategreen{-1}{2}
		\MYcircle{-0.5}{-0.5}
		\MYcircle{-0.5}{0.5}
		\MYcircle{-1.5}{1.5}
		\MYcircle{-1.5}{2.5}
		\MYcircle{-0.5}{-1.5}
  \Text[x=-0,y=-2]{$ $}
		\end{tikzpicture}.
	\end{eqnarray}

It can be shown, using a simplification procedure similar to that in Sec.~\ref{subsec:STCF}, that this condition leads to the exact solvability of two point spatial-temporal correlation functions in the region $-t/3<x<0$. This is interesting since it gives us a 2D space-time region where the spatial-temporal correlation functions are solvable and non-vanishing, while all dual-unitary circuits and their $\mathfrak{L}_2$ generalization have vanishing spatial-temporal correlations except on a few lines. Unfortunately, to date, we have not yet been able to find a nontrivial unitary gate satisfying this condition, due to the computational difficulty of solving this equation. We leave this possibility as an open direction for the future.

\section*{Acknowledgements}
We thank Ignacio Cirac, Katja Klobas, Bruno Bertini and Alessandro Foligno  for fruitful discussions.
PK is supported by the Alexander von Humboldt Foundation. ZW is supported by the Munich Quantum Valley~(MQV), which is supported by the Bavarian state government with funds from the Hightech Agenda Bayern Plus.
\vspace{.5cm}

\appendix
\section{A more general parametrization of dual-unitarity and hierarchy gates based on finite groups} \label{Appendix:general}
In this appendix we give a further generalization of the parametrization of dual-unitarity and hierarchy gates presented in Sec.~\ref{sec:DU}, based on projective representations of finite groups. 
Let $G$ be a finite group and let $\Gamma$ be an irreducible, unitary  finite dimensional~(with dimension $D$) projective representation of $G$, i.e. $\Gamma_a^\dagger=\Gamma_{a^{-1}},\forall a\in G$ and 
\begin{equation}\label{eq:def_projRep}
    \Gamma_a\Gamma_b=\phi(a,b) \Gamma_{ab},~~\forall a,b\in G,
\end{equation}
where $\phi(a,b)$ is a 2-cocycle of $G$, i.e.
\begin{equation}
    \phi(a,b)\phi(ab,c)=\phi(b,c)\phi(a,bc), ~~\forall a,b,c\in G.
\end{equation}
We further require that $\Gamma$ satisfies a trace condition 
\begin{equation}\label{eq:PSGtracecondition}
    \mathrm{Tr}[\Gamma_a]=D \delta_{a,e},~~\forall a\in G,
\end{equation}
where $e$ is the unit of $G$. The irreducibility of $\Gamma$ implies that  $\{\Gamma_a\}_{a\in G}$ spans the matrix algebra $M_D(\mathbb{C})$, therefore, $\{\Gamma_a\}_{a\in G}$ is an orthonormal basis of  $M_D(\mathbb{C})$ with trace normalization $\mathrm{Tr}[\Gamma_a^\dagger \Gamma_b]=D\delta_{ab},\forall a,b\in G$. 

We use Eq.~\eqref{eq:2quditUgeneralform} again for the parametrization of 2-qudit unitary gate $u$, but now $u_0$ is defined as 
\begin{equation}
	u_0=\sum_{a\in G} \theta_{a}\ket{\psi_{a}}\bra{\psi_{a}},
	\label{eq:PRGParametrization}
\end{equation}
where $\{\theta_{a}\}_{a\in G}$ is a collection of $\mathrm{U}(1)$ phases, and $\{\ket{\psi_{a}}\}_{a\in G}$ is an orthonormal basis for the 2-qudit Hilbert space $(\mathbb{C}^D)^2$ defined as
\begin{equation}
	\ket{\psi_{a}}\equiv \frac{1}{\sqrt{D}}\sum_{0\leq i,j\leq D-1}(\Gamma_a)^*_{ij}\ket{i}\otimes\ket{j}.
	\label{eq:psi_a}
\end{equation}

We proceed to investigate conditions on the parameters $\{\theta_{a}\}_{a\in G}$ for $u$ to be a dual-unitary gate. After a space-time reshuffling of indices defined in Eq.~\eqref{eq:tildeqgate}, we have $\tilde{u}= (v_4^T\otimes v_2)\tilde{u}_0(v_3\otimes v_1^T)$, where
\begin{equation}
	\tilde{u}_0=\frac{1}{D}\sum_{a\in G} \theta_{a}\Gamma_{a}\otimes\Gamma^*_{a}.
\end{equation}
Then the unitarity condition~\eqref{eq:algebric_dual_unitary} on $\tilde{u}$ is equivalent to 
\begin{equation}\label{eq:simplifyunitarityPRG}
	\sum_{a,b\in G} \theta_{a}\theta^*_{b}\Gamma_{a}\Gamma^\dagger_{b}\otimes\Gamma^*_{a}\Gamma^T_{b}=D^2 \Gamma_{e}\otimes\Gamma_{e}.
\end{equation}
We simplify Eq.~\eqref{eq:simplifyunitarityPRG} further with Eq.~\eqref{eq:def_projRep} and, using the fact that $\{\Gamma_{a}\}_{a\in G}$ forms a basis of the matrix algebra $M_D(\mathbb{C})$, we obtain 
\begin{equation}\label{eq:DUcondition_theta_PRG}
	\sum_{b\in G} \theta_{ab} \theta^*_{b} =D^2\delta_{a,e},~~ \forall a\in G. 
\end{equation}
Note that Eq.~\eqref{eq:DUcondition_theta_PRG} does not depend on the cocycle $\phi(a,b)$, only on the structure of the group $G$. Nevertheless, not all projective representations satisfy the trace condition~\eqref{eq:PSGtracecondition}, and the 2-cocycle has to be chosen carefully to allow for such a projective representation.  

As a specific example, consider the Abelian group $G=\mathbb{Z}_k^{\times 2n}$, where group elements are denoted as $a=(a_0,a_1,\ldots,a_{2n-1}), 0\leq a_j\leq k-1$, with multiplication rule $(a_0,\ldots,a_{2n-1})(b_0,\ldots,b_{2n-1})=(a_0+b_0,\ldots,a_{2n-1}+b_{2n-1})$~(where all additions are modulo $k$), and the 2-cocycle is given by 
\begin{equation}
    \phi(a,b)=\omega^{-\sum_{0\leq i<j\leq 2n-1}b_i a_j},
\end{equation}
where $\omega$ is a $k$-th root of unity. The projective representation $\Gamma$ is defined as 
\begin{equation}
    \Gamma_a=\gamma_{0}^{a_0}\gamma_1^{a_1}\ldots\gamma_{2n-1}^{a_{2n-1}},
\end{equation}
where $\gamma_0,\gamma_1,\ldots \gamma_{2n-1}$ are $D\times D$ matrices with $D=k^n$ satisfying 
\begin{eqnarray}
    \gamma_j^k&=&1,~~0\leq j\leq 2n-1,\nonumber\\
    \gamma_i\gamma_j&=&\omega \gamma_j\gamma_j,~~ 0\leq i<j\leq 2n-1.
\end{eqnarray}
The parametrization presented in Sec.~\ref{sec:DU} corresponds to the special case $n=1$.

\section{The parametrization of the Hierarchical gates in higher dimensions}\label{sec:appendixA}

In this appendix, we provide further details about deriving Eq. (\ref{eq:Clifford_parametrization_2nd_result}). In particular, inserting Eqs. (\ref{eq:2quditUgeneralform}) and (\ref{eq:CliffordParametrization}) into the first equation of (\ref{eq2:bottomtotop})
with $v_{3}=v_{4}=I_{D}$, we obtain
\begin{widetext}
\begin{align}
 & \sum_{a,b,c,d}\theta_{p_{b},q_{b}}^{*}\theta_{p_{a},q_{a}}\theta_{p_{d},q_{d}}^{*}\theta_{p_{c},q_{c}}\tau_{p_{b},q_{b}}^{\dagger}\tau_{p_{a},q_{a}}\otimes\tau_{p_{d},q_{d}}^{\dagger}v_{1}^{*}\tau_{p_{b},q_{b}}^{T}\tau_{p_{a},q_{a}}^{*}v_{1}^{T}\tau_{p_{c},q_{c}}\otimes\tau_{p_{d},q_{d}}^{T}\tau_{p_{c},q_{c}}^{*}\nonumber \\
=D^{2}\sum_{c,d} & \theta_{p_{d},q_{d}}^{*}\theta_{p_{c},q_{c}}I_{D}\otimes\tau_{p_{d},q_{d}}^{\dagger}\tau_{p_{c},q_{c}}\otimes\tau_{p_{d},q_{d}}^{T}\tau_{p_{c},q_{c}}^{*}.
\end{align}
With the help of Eq. (\ref{eq:relations_tau_pq}), this can be simplified to 

\begin{equation}
\begin{aligned} & \sum_{a,b,c,d}\omega^{q_{d}p_{c}}\theta_{p_{b},q_{b}}^{*}\theta_{p_{a},q_{a}}\theta_{p_{d},q_{d}}^{*}\theta_{p_{c},q_{c}}\tau_{p_{a}-p_{b},q_{a}-q_{b}}\otimes\tau_{-p_{d},-q_{d}}v_{1}^{*}\tau_{p_{a}-p_{b},q_{b}-q_{a}}v_{1}^{T}\tau_{p_{c},q_{c}}\otimes\tau_{p_{c}-p_{d},q_{d}-q_{c}}\\
= & D^{2}\sum_{c,d}\theta_{p_{d},q_{d}}^{*}\theta_{p_{c},q_{c}}I_{D}\otimes\tau_{p_{c}-p_{d},q_{c}-q_{d}}\otimes\tau_{p_{c}-p_{d},q_{d}-q_{c}}.
\end{aligned}
\end{equation}
We can relabel the dummy variables $p_{a}=p_{b}+k,q_{a}=q_{b}+l,p_{c}=p_{d}+s,q_{c}=q_{d}+t$
so that the independent Clifford group matrix is separated as
\begin{equation}
\begin{aligned} & \sum_{k,l}\left(\sum_{b}\theta_{p_{b},q_{b}}^{*}\theta_{p_{b}+k,q_{b}+l}\right)\tau_{k,l}\otimes\sum_{s,t}\left(\sum_{d}\theta_{p_{d},q_{d}}^{*}\theta_{p_{d}+s,q_{d}+t}\tau_{p_{d},q_{d}}^{\dagger}v_{1}^{*}\tau_{k,-l}v_{1}^{T}\tau_{p_{d},q_{d}}\right)\tau_{s,t}\otimes\tau_{s,-t}\\
= & D^{2}\sum_{s,t}\sum_{d}\theta_{p_{d},q_{d}}^{*}\theta_{p_{d}+s,q_{d}+t}I_{D}\otimes\tau_{s,t}\otimes\tau_{s,-t},
\end{aligned}
\label{eq:Clifford_equation_after_simpli}
\end{equation}
where we have used the equality $\tau_{p_{c},q_{c}}=\tau_{p_{d},q_{d}}\tau_{s,t}\omega^{-q_{d}s}$. 

Note that Eq. (\ref{eq:Clifford_equation_after_simpli}) is automatically
satisfied for $(k,l)=(0,0)$. Therefore, for $\forall(k,l)\neq(0,0)$,
the left hand side of Eq. (\ref{eq:Clifford_equation_after_simpli})
must vanish, namely, either $\sum_{b}\theta_{p_{b},q_{b}}^{*}\theta_{p_{b}+k,q_{b}+l}=0$
or $\sum_{d}\theta_{p_{d},q_{d}}^{*}\theta_{p_{d}+s,q_{d}+t}\tau_{p_{d},q_{d}}^{\dagger}v_{1}^{*}\tau_{k,-l}v_{1}^{T}\tau_{p_{d},q_{d}}=0$,
which gives the first line of Eq. (\ref{eq:Clifford_parametrization_2nd_result}) in the main text. The second
line of Eq. (\ref{eq:Clifford_parametrization_2nd_result}) containing $v_{2}$ simplifies in a similar way using the second equation of (\ref{eq2:bottomtotop}).

The derivation of the $\mathfrak{L}_3$ condition follows
 exactly same procedure as above but may involve lots of tedious
calculations. Here we just skip these details and give the result
directly. With the parametrization method using Clifford group, Eq.
(\ref{eq:3rdHierarchydefinition}) is equivalent to
\begin{align}
 & \left(\sum_{p_{b},q_{b}}\theta_{p_{b}+s_{a},q_{b}+t_{a}}\theta_{p_{b},q_{b}}^{*}\right)\left(\sum_{p_{d},q_{d}}\theta_{p_{d}+s_{c},q_{d}+t_{c}}\theta_{p_{d},q_{d}}^{*}\tau_{p_{d},q_{d}}^{\dagger}v_{1}^{*}\tau_{s_{a},-t_{a}}v_{1}^{T}\tau_{p_{d},q_{d}}\right)\nonumber \\
\times & \left(\sum_{p_{f},q_{f}}\theta_{p_{f}+s_{e},q_{f}+t_{e}}\theta_{p_{f},q_{f}}^{*}\tau_{p_{f},q_{f}}^{\dagger}v_{1}^{*}\tau_{s_{c},-t_{c}}v_{1}^{T}\tau_{p_{f},q_{f}}\right)=0\ \ \mathrm{for}\ \forall(s_{a},t_{a})\neq(0,0) \ .
\end{align}
\end{widetext}


\section{Examples of $\overline{\mathfrak{L}}_2$ Hierarchical circuits}\label{appen:detailed_of_example}

\subsection{Exhaustive parametrization for qubits}
\label{app:qubitsDetails}
In the qubit case, the full parametrization of a two-qubit gate Eq. (\ref{eq:parameter2qubit}) is used with $v_3=v_4=I$. The Hierarchical condition Eq. (\ref{eq2:bottomtotop}) or Eq. (\ref{eq:2ndEq}) can be transformed into a series of algebraic equations of $J_x,J_y,J_z,r_i,\theta_i,\phi_i$ with $v_i=\exp\{ir_i(\cos\theta_i\sigma_z+\sin\theta_i\cos\phi_i\sigma_x+\sin\theta_i\sin\phi_i\sigma_y)\}$, $i\in\{1,2\}$. However, these equations are very complicated and virtually impossible to be analytically solved. 

Our strategy is to combine the analytical analysis with the numerical help of Mathematica. Those algebraic equations can be represented as a cost function $f \geq 0$, designed to be minimized. The equations are solved for $f=0$.
The minimization is done by the Mathematica function NMinimize and we apply different constraints to it. For example, we require that all three $J$s are nonvanishing, i.e. not equal to $n\pi/2,n\in\mathbb{Z}$, and also avoid the dual unitary case. Extensive numerical experiments very strongly suggest that it is impossible to achieve a minimum value close to $0$ under these constraints, leading to the conclusion that a vanishing $J$ is necessary to satisfy the condition of $\overline{\mathfrak{L}}_2$. Consequently, we proceed by setting $J_x=0$ in the subsequent analysis. This process of simplification and analysis continues until the resultant algebraic equations become tractable for manual examination.

\subsection{Non-exhaustive parametrization for qudits with $D>2$}
\label{app:quditsDetails}
Here we provide some more details regarding examples with $D>2$ mentioned in the main text.
In particular, here we explain how to derive the simplified conditions that $v_1$ and $v_2$ in Eq. (\ref{eq:Clifford_parametrization_2nd_result}) need to satisfy. 
These simplified conditions help obtain $v_1$ and $v_2$. In some cases, one can guess good candidates, and in other cases, these simplified conditions help in finding examples numerically. 
However, an explicit construction of all possible $v_1$ and $v_2$ is still lacking.

Given that the Clifford group matrix $\tau_{r,m}$ is complete, the expression $v_1^*\tau_{k,-l}v_1^T$ from the first line of Eq. (\ref{eq:Clifford_parametrization_2nd_result}) can be decomposed into
\begin{equation}
v_1^*\tau_{k,-l}v_1^T=\sum_{0\leq r,m\leq D-1}\alpha_{r,m}\tau_{r,m},
\end{equation}
where $\alpha$s are the coefficients. By substituting this decomposition 
into the first line of Eq. (\ref{eq:Clifford_parametrization_2nd_result}), and following a direct calculation of the Clifford group matrix, we arrive at
\begin{widetext}

\begin{equation}
\left(\sum_{b}\theta_{p_{b},q_{b}}^{*}\theta_{p_{b}+k,q_{b}+l}\right)\left(\sum_{r,m}\alpha_{r,m}\tau_{r,m}\sum_{d}\theta_{p_{d},q_{d}}^{*}\theta_{p_{d}+s,q_{d}+t}\omega^{p_dm-q_dr}\right)=0,\; \forall\{k,l\}\neq\{0,0\},\;\forall\{s,t\}\neq\{0,0\}.\label{Eq:simplified_a_little}
\end{equation}
\end{widetext}
We can look into the two terms separately. If the first term vanishes for all of $\{k,l\}\neq\{0,0\}$, we just obtain the $\mathfrak{L}_1$ condition Eq. (\ref{eq:DUcondition_theta}). Thus, a $\overline{\mathfrak{L}}_2$ circuit would necessarily require that for some $\{k,l\}\neq\{0,0\}$, the first term is non-vanishing. The set of these $\{k,l\}$ can be directly determined with the specific form of $\theta_{p,q}$. The second term must vanish for this set of $\{k,l\}$s to satisfy Eq. (\ref{Eq:simplified_a_little}). 

Since $\tau_{r,m}$ is a complete basis in the matrix space, the necessary and sufficient condition for the vanishing of the second term can be expressed as: for every $\{r,m\}$, either $\alpha_{r,m}=0$ or
\begin{equation}
\sum_{d}\theta_{p_{d},q_{d}}^{*}\theta_{p_{d}+s,q_{d}+t}\omega^{p_dm-q_dr}=0,\;\forall\{s,t\}\neq\{0,0\}.\label{Eq:determining_rm}
\end{equation} The latter one can also be directly calculated by substituting with the specific form of $\theta_{p,q}$. If it does not vanish for some $\{r,m\}$s, we obtain the condition that $\alpha_{r,m}$ must vanish for those pairs. Since $\alpha_{r,m}$ is related to $v_1$, this serves as a necessary and sufficient condition $v_1$ should satisfy. 

In conclusion, $v_1$ is implicitly defined by its conjugation action on those $\tau_{k,-l}$, with $\{k,l\}$ determined from the first term in Eq. (\ref{Eq:simplified_a_little}), such that $v_1^*\tau_{k,-l}v_1^T$ has zero overlap with certain $\tau_{r,m}$, with $\{r,m\}$ determined from Eq. (\ref{Eq:determining_rm}).

We mention that this condition is more convenient than the original $\overline{\mathfrak{L}}_2$ condition since we use the property of the Clifford group matrix. To construct $v_1$ satisfying the condition, one can either guess directly or use some ansatz of $v_1$ to solve the condition. The same procedure works for $v_2$.


For the class given by the Eq.~\eqref{eq:simplest_2nd_Hierarchy_case} of the main text, the above procedure results in the implicit equations for $v_i$ 
\begin{equation}
\text{Tr}\ \ v_i^*\tau_{k,l}v_i^T\tau_{r,v}=0,
\end{equation}
where $i\in\{1,2\}$ and the equation holds for $\forall k,l=\mathrm{even},(k,l)\neq(0,0)\; \text{and}\;r,v \in \{0,D/2\}$. A naive example of $v_i$ is just the identity. In $D=6$ we can use the ansatz $v_i=\sigma_x\otimes\kappa_i$ where $\sigma_x$ is the Pauli-x operator. We numerically find that $v_i$ satisfies the above condition for almost any $\kappa_i\in\mathbb{SU}(3)$.

For the class given by Eq. (\ref{eq:second_example}), the condition can be expressed as $\text{Tr} \; v_i^*\tau_{0,l}v_i^T\tau_{0,v}=0\,\text{for}\; \forall v,l\neq0$ where $i\in\{1,2\}$. A simple example that satisfies the condition is the generalized Hamamard gate $H$ in higher dimension defined as $H\sigma H^\dagger=\tau$. 

\section{Proof of Eq.~\eqref{eq:2ndfig}}\label{app:proof2nd}
In this subsection, we prove that the left equality in Eq. (\ref{eq2:bottomtotop}) along with unitarity implies the left equality in Eq. (\ref{eq:2ndfig}). The same argument holds for the right one.
To prove this implication, we introduce the definitions \begin{equation}
A=
\begin{tikzpicture}[baseline=(current  bounding  box.center), scale=0.5]
\Wgategreen{-0.5}{-0.5}
\Wgategreen{0.5}{0.5}
\MYcircle{1.05}{1.05}
\MYcircle{1.05}{-0.05}
\MYcircle{0.05}{-1.05}
\end{tikzpicture}
-
\begin{tikzpicture}[baseline=(current  bounding  box.center), scale=0.5]
\Wgategreen{-0.5}{-0.5}
\MYcircle{1.05}{1.05}
\MYcircle{0.05}{0.05}
\MYcircle{0.05}{-1.05}
\draw[thick] (0,1.05)--(1,1.05);
\end{tikzpicture}
,
B=
\begin{tikzpicture}[baseline=(current  bounding  box.center), scale=0.5]
\Wgategreen{-0.5}{-0.5}
\Wgategreen{0.5}{0.5}
\MYcircle{-0.05}{1.05}
\MYcircle{-1.05}{-0.05}
\MYcircle{-1.05}{-1.05}
\end{tikzpicture}
-
\begin{tikzpicture}[baseline=(current  bounding  box.center), scale=0.5]
\Wgategreen{-0.5}{-0.5}
\MYcircle{-0.05}{1.05}
\MYcircle{-1.05}{0.05}
\MYcircle{-1.05}{-1.05}
\draw[thick] (0,1.05)--(1,1.05);
\end{tikzpicture}
.
\end{equation}

By proving $\mathrm{Tr}A^{\dagger}A=\mathrm{Tr}B^{\dagger}B$, we establish that if $A$ vanishes, then $B$ also vanishes. This
can be proved graphically. To this end, we introduce a four-folded
gate given by$
\begin{tikzpicture}[baseline=(current  bounding  box.center), scale=0.7]
\Wgatedagger{0}{0}
\end{tikzpicture}
=
\begin{tikzpicture}[baseline=(current  bounding  box.center), scale=0.7]
\Wgatered{-0.3}{-0.3}
\Wgateblue{-0.1}{-0.1}
\Wgatered{0.1}{0.1}
\Wgateblue{0.3}{0.3}
\end{tikzpicture}
.
$
 The first term of $\mathrm{Tr}A^{\dagger}A$ is

\begin{equation}
\mathrm{Tr}
\left(
\begin{tikzpicture}[baseline=(current  bounding  box.center), scale=0.5]
\Wgategreen{-0.5}{-0.5}
\Wgategreen{0.5}{0.5}
\MYcircle{1.05}{1.05}
\MYcircle{1.05}{-0.05}
\MYcircle{0.05}{-1.05}
\end{tikzpicture}
\right)^{\dagger}
\left(
\begin{tikzpicture}[baseline=(current  bounding  box.center), scale=0.5]
\Wgategreen{-0.5}{-0.5}
\Wgategreen{0.5}{0.5}
\MYcircle{1.05}{1.05}
\MYcircle{1.05}{-0.05}
\MYcircle{0.05}{-1.05}
\end{tikzpicture}
\right)
=
\begin{tikzpicture}[baseline=(current  bounding  box.center), scale=0.5]
\Wgatedagger{-0.5}{-0.5}
\Wgatedagger{0.5}{0.5}
\MYcircle{1.05}{1.05}
\MYcircle{1.05}{-0.05}
\MYcircle{0.05}{-1.05}
\MYsquare{-0.05}{1.05}
\MYsquare{-1.05}{-0.05}
\MYsquare{-1.05}{-1.05}
\end{tikzpicture}
.
\end{equation}
The contraction
\begin{tikzpicture}[baseline=(current  bounding  box.center), scale=0.7]
\MYcircle{0.25}{0.25}
\draw[very thick] (-0.25,-0.25)--(0.20,0.20);
\end{tikzpicture} represents a contraction between the same leg from the first and
second layer as well as the third and fourth layer.
Similarly, the contraction
\begin{tikzpicture}[baseline=(current  bounding  box.center), scale=0.7]
\MYsquare{0.25}{0.25}
\draw[very thick] (-0.25,-0.25)--(0.20,0.20);
\end{tikzpicture} represents a contraction between the first and fourth layer as well
as second and third layer. It is worth noting that the result remains the same if we exchange
the second and fourth layers while simultaneously exchange 
\begin{tikzpicture}[baseline=(current  bounding  box.center), scale=0.7]
\MYcircle{0.25}{0.25}
\draw[very thick] (-0.25,-0.25)--(0.20,0.20);
\end{tikzpicture} and \begin{tikzpicture}[baseline=(current  bounding  box.center), scale=0.7]
\MYsquare{0.25}{0.25}
\draw[very thick] (-0.25,-0.25)--(0.20,0.20);
\end{tikzpicture}. 
Therefore, we arrive at

\begin{equation}
\mathrm{Tr}
\left(
\begin{tikzpicture}[baseline=(current  bounding  box.center), scale=0.5]
\Wgategreen{-0.5}{-0.5}
\Wgategreen{0.5}{0.5}
\MYcircle{1.05}{1.05}
\MYcircle{1.05}{-0.05}
\MYcircle{0.05}{-1.05}
\end{tikzpicture}
\right)^{\dagger}
\left(
\begin{tikzpicture}[baseline=(current  bounding  box.center), scale=0.5]
\Wgategreen{-0.5}{-0.5}
\Wgategreen{0.5}{0.5}
\MYcircle{1.05}{1.05}
\MYcircle{1.05}{-0.05}
\MYcircle{0.05}{-1.05}
\end{tikzpicture}
\right)
=
\begin{tikzpicture}[baseline=(current  bounding  box.center), scale=0.5]
\Wgatedagger{-0.5}{-0.5}
\Wgatedagger{0.5}{0.5}
\MYcircle{1.05}{1.05}
\MYcircle{1.05}{-0.05}
\MYcircle{0.05}{-1.05}
\MYsquare{-0.05}{1.05}
\MYsquare{-1.05}{-0.05}
\MYsquare{-1.05}{-1.05}
\end{tikzpicture}
=
\begin{tikzpicture}[baseline=(current  bounding  box.center), scale=0.5]
\Wgatedagger{-0.5}{-0.5}
\Wgatedagger{0.5}{0.5}
\MYsquare{1}{1}
\MYsquare{1}{0}
\MYsquare{0}{-1}
\MYcircle{0}{1}
\MYcircle{-1}{0}
\MYcircle{-1}{-1}
\end{tikzpicture}
.
\end{equation}

The R.H.S. just corresponds to the first term in $\mathrm{Tr}B^{\dagger}B$. By performing similar calculations,
we can establish the agreement between each term of $\mathrm{Tr}A^{\dagger}A$ and
$\mathrm{Tr}B^{\dagger}B$.

The fact that a Hierarchical condition along with unitarity implies that its $180^\circ$ rotated version of the condition can be straightforwardly extended to higher-level Hierarchical circuits. The derivation presented here applies almost imediately.

\section{Numerical values of the parameters of the gates} \label{app:parameters}

In Fig. \ref{Correlation_func_sup2}(b) and Fig. \ref{quench_correlation},
we choose single site operators as $v_{3}=v_{4}=I$, $v_{1}=v_{2}=v$.
The parameters for $v=e^{ir(\cos\theta\sigma_{z}+\sin\theta\cos\phi\sigma_{x}+\sin\theta\sin\phi\sigma_{y})}$ are listed in Table  \ref{tab:my_label}.
\begin{table}[!htbp]
    \centering
\begin{tabular}{|c|c|c|}
\hline 
$r$ & $\theta$ & $\phi$\tabularnewline
\hline 
\hline 
$1.24056$ & $0.84429$ & $-0.4764$\tabularnewline
\hline 
\end{tabular}
\ \ \
\begin{tabular}{|c|c|c|}
\hline 
$r$ & $\theta$ & $\phi$\tabularnewline
\hline 
\hline 
$1$ & $0.99788$ & $\text{3}$\tabularnewline
\hline 
\end{tabular} \ \ \ 
\begin{tabular}{|c|c|c|}
\hline 
$r$ & $\theta$ & $\phi$\tabularnewline
\hline 
\hline 
$\frac{\pi}{4}$ & $\frac{\pi}{2}$ & $0$\tabularnewline
\hline 
\end{tabular}
\caption{The left set of parameters is the one for Fig. \ref{Correlation_func_sup2}(b) and the blue line in Fig. \ref{quench_correlation}. The middle set corresponds to the green line in Fig. \ref{quench_correlation} and the right set corresponds to the red line in Fig. \ref{quench_correlation}.}
\label{tab:my_label}
\end{table}
\vspace{0.6cm}
\bibliographystyle{quantum}
\bibliography{bibliography}
\end{document}